\documentclass[prb,reprint,a4paper,groupedaddress,showpacs,citeautoscript]{revtex4-1}

\pdfoutput=1
\usepackage[charter]{mathdesign}
\usepackage{amsmath}

\usepackage[pdftex]{graphicx}
\usepackage{microtype}
\usepackage{bm}\let\vec\bm
\usepackage[innercaption]{sidecap}

\usepackage[svgnames]{xcolor}
\usepackage[
    colorlinks=True,linkcolor=DarkRed,citecolor=ForestGreen,urlcolor=MediumBlue,
    pdfstartview=FitH,bookmarks=False,pdfpagemode=UseNone
]{hyperref}

\begin{document}

\title{\boldmath Strong-coupling analysis of scanning tunneling spectra in
Bi$_2$Sr$_2$Ca$_2$Cu$_3$O$_{10+\delta}$}

\author{C. Berthod}
\author{Y. Fasano}
\altaffiliation[Present address: ]{Low Temperatures Lab, Centro At{\'o}mico
Bariloche, Argentina.}
\author{I. Maggio-Aprile}
\author{A. Piriou}
\author{E. Giannini}
\author{G. Levy de Castro}
\altaffiliation[Present address: ]{Department of Physics and Astronomy,
University of British Columbia, Vancouver, British Columbia V6T 1Z1, Canada.}
\author{{\O}. Fischer}
\affiliation{DPMC-MaNEP, Universit{\'e} de Gen{\`e}ve, 24 quai Ernest-Ansermet,
1211 Gen{\`e}ve 4, Switzerland}

\date{11 March 2013}

\begin{abstract}

We study a series of spectra measured in the superconducting state of
optimally-doped Bi$_2$Sr$_2$Ca$_2$Cu$_3$O$_{10+\delta}$ (Bi-2223) by scanning
tunneling spectroscopy. Each spectrum, as well as the average of spectra
presenting the same gap, is fitted using a strong-coupling model taking into
account the band structure, the BCS gap, and the interaction of electrons with
the spin resonance. After describing our measurements and the main
characteristics of the strong-coupling model, we report the whole set of
parameters determined from the fits, and we discuss trends as a function of the
gap magnitude. We also simulate angle-resolved photoemission spectra, and
compare with recent experimental results.

\end{abstract}

\maketitle

\section{Introduction}

The two main single-electron spectroscopies, angle-resolved photoemission
\cite{Damascelli-2003} (ARPES) and scanning tunneling microscopy
\cite{Fischer-2007} (STM), have considerably improved during the last decades,
mainly motivated by the quest for reliable data in the study of cuprate
high-temperature superconductors. A growing body of high-quality spectroscopic
data is now available for the cuprates, especially for compounds of the bismuth
family, which offer clean surfaces. These refined experiments on high-quality
crystals may be able to deliver the intrinsic line shape of the one-electron
spectra. Nonetheless, very few studies have undertaken a detailed line-shape
study of the various spectral features by means of a microscopic model.
Following pioneering studies \cite{Renner-1995, Mallet-1996}, the STM data
analysis has remained mostly qualitative, or based on phenomenological
approaches. In a few cases, a BCS $d$-wave model including a realistic band
structure \cite{Piriou-2011}, and/or a phenomenological scattering rate
\cite{Alldredge-2008}, turned out to be appropriate. Such cases are the
exception rather than the rule: these models will not capture, in particular,
the ``dip'' feature ubiquitously present at energies above the superconducting
gap. By analogy with phonon-related effects in classical superconductors
\cite{Giaever-1962, McMillan-1965}, the dip, also recently observed in pnictide
superconductors \cite{Fasano-2010, Chi-2012}, is generally attributed to a
collective mode. In order to analyze the phenomenon, an extension of the
Eliashberg formalism to $d$-wave superconductors has been proposed
\cite{Zasadzinski-2003}. Different interpretations, based on phonons
\cite{Devereaux-2004, *Johnston-2010, *Johnston-2010b, Nieminen-2012}, an
energy-dependent gap function \cite{Cren-2000a, *Sacks-2006}, gap
inhomogeneities \cite{Fang-2006}, or a charge-density wave order
\cite{Gabovich-2007}, have also been put forward.

In the superconducting state, the cuprates present a low-energy magnetic
excitation known as the spin resonance. This excitation is observed below the
superconducting critical temperature, $T_c$, by inelastic neutron scattering, as
a strong enhancement of the spin susceptibility around the anti-ferromagnetic
vector $\vec{Q}=(\pi/a,\pi/a)$. First discovered in YBa$_2$Cu$_3$O$_{6+x}$
(Y-123) \cite{Rossat-Mignod-1991, Mook-1993} at an energy $\Omega_s=41$~meV, it
was later observed in most cuprates, including the single-layer compounds
HgBa$_2$CuO$_{4+\delta}$ \cite{Yu-2010} and Tl$_2$Ba$_2$CuO$_{6+\delta}$
\cite{He-2002, *Keimer-2002}, the two-layer Bi$_2$Sr$_2$CaCu$_2$O$_{8+\delta}$
(Bi-2212), \cite{Fong-1999, *Capogna-2007, *Fauque-2007, Mesot-2000, Xu-2009}
and the electron-doped material Pr$_{0.88}$LaCe$_{0.12}$CuO$_{4-\delta}$
\cite{Wilson-2006}. The $(\pi,\pi)$ resonance energy ranges from 10 to 60~meV,
roughly correlated with $T_c$ as $\Omega_s\approx5.3k_{\text{B}}T_c$
\cite{Sidis-2004}. It is found in both underdoped and overdoped materials
\cite{He-2001}, and was also detected above $T_c$ in Y-123 \cite{Stock-2004}.
The resonance has been interpreted as a spin-1 exciton, bound below the
continuum of electron-hole excitations, gapped by the superconducting pairing
\cite{*[{See, e.g., }] [{, and references therein.}] Eschrig-2006}. One of the
open questions concerns the role played by spin fluctuations, and particularly
by the $(\pi,\pi)$ resonance, in the pairing phenomenon \cite{Anderson-2007,
Maier-2008}.

Being related to pair formation or not, the $(\pi,\pi)$ resonance is a
collective spin excitation, which must somehow interact with the charge degrees
of freedom and induce renormalization and damping of the Bogoliubov
quasiparticles in the superconducting state. This interaction affects the
one-electron spectra and is observable in the single-electron spectroscopies.
Although the strength of this interaction has been a matter of controversy
\cite{Kee-2002, Abanov-2002}, there is evidence that peculiar signatures
observed in photoemission \cite{Norman-1997, *Campuzano-1999, *Kaminski-2001,
Kim-2003, *Borisenko-2003, *Zabolotnyy-2006, *Borisenko-2006, Gromko-2003,
Sato-2003}, tunneling \cite{Zasadzinski-2001, Hoogenboom-2003b, Levy-2008,
Jenkins-2009, Niestemski-2007}, and optical conductivity \cite{Carbotte-1999,
*Tu-2002, *Hwang-2004, *Yang-2009, vanHeumen-2009a, *vanHeumen-2009b} result
from this interaction. Yet, a firm consensus has not been reached: optical
phonons often exist in the cuprates at similar energies, and distinguishing the
effects of the two kinds of excitations has proven difficult. The spin
resonance, being localized near $(\pi,\pi)$, leads to an anisotropic scattering
rate and a strong dip \cite{Levy-2008}. But similar effects can be induced by
phonons, provided that the electron-phonon coupling is strongly anisotropic
\cite{Devereaux-2004, *Johnston-2010, *Johnston-2010b}. A possible way of
determining the origin of the dip feature is to study its evolution with doping,
to be compared with the doping dependence of the spin resonance and phonons,
both directly measured by neutron scattering.

In a $d$-wave superconductor, one of the signatures of the coupling to a
collective mode is a \emph{minimum}, so-called dip, in the electron density of
states (DOS), occurring at an energy $E_d$, which is separated from the energy
$\Delta_p$ (of the coherence peak) by the mode energy \cite{Eschrig-2000,
*Eschrig-2003}. In contrast, for $s$-wave superconductors, the signature is a
\emph{change of curvature} of the DOS, leading to a peak in the DOS derivative
\cite{Schrieffer-1963, Berthod-2010}. Scanning tunneling microscopy/spectroscopy
(STM/STS) allows one to measure the gap in the excitation spectrum, as well as
the dip, with sub-meV and atomic resolutions, and to track their spatial
variations in inhomogeneous materials \cite{Jenkins-2009}. It is therefore an
ideal tool to investigate the properties of the dip and the relationship between
the gap and the resonance energy. The three-layer compound
Bi$_2$Sr$_2$Ca$_2$Cu$_3$O$_{10+\delta}$ (Bi-2223) is well suited for such
studies. It can be cleaved and offers atomically flat surfaces for STM
investigations. Bi-2223 has the highest optimal $T_c$ of the bismuth family,
111~K, a gap in the 30--60~meV range, and a very strong dip as revealed by
tunneling \cite{Kugler-2006} and photoemission \cite{Sato-2002, Feng-2002,
Matsui-2003a}.

In this paper, we study Bi-2223 STS spectra by means of least-squares fits to a
strong-coupling model consisting of the Van Hove singularity (VHS) associated
with the saddle point of a two-dimensional tight-binding band, a $d$-wave BCS
gap, and a coupling to the $(\pi,\pi)$ resonance \cite{Eschrig-2000}. The
results of similar studies were reported previously \cite{Levy-2008,
Jenkins-2009}. Here, we describe our fits in detail, we fit average as well as
local spectra, we provide and discuss the complete set of model parameters, and
we use these parameters to simulate ARPES data. The motivation for performing
direct \emph{fits} to cuprate STS data is twofold. The first aim is to
demonstrate that in spite of its simplicity, the model captures quantitatively
the main characteristics of the data for optimally doped Bi-2223: a V-shaped gap
at low energy, tall coherence peaks and very pronounced dips, both significantly
electron-hole asymmetric. Second, the quality of these fits provides further
evidence that the STS tunneling conductance measures the full \emph{electron}
local DOS (LDOS) \cite{Piriou-2011}, rather than an effective quasiparticle DOS
deprived of band-structure and self-energy effects \cite{Anderson-2006,
*Anderson-2011}.

In Sec.~\ref{sec:experiment}, we describe the growth and characterization of the
samples, present the measurement method, and discuss the main features of the
spectra. Section~\ref{sec:theory} is dedicated to the model. We use different
conventions than Ref.~\onlinecite{Eschrig-2000} for the model parameters. For
definiteness, we describe and discuss our model in detail. We also explain the
fitting method. In Sec.~\ref{sec:results}, we present our results and the trends
in fitted parameters. We discuss the values of the most important parameters in
Sec.~\ref{sec:discussion}, compare with values obtained using other experimental
techniques, and present simulations of ARPES intensities. Finally,
Sec.~\ref{sec:conclusion} is a summary of the results and implications of the
present study.

\section{Experimental details}
\label{sec:experiment}

\subsection{Sample growth and characterization, STM measurements}

The Bi-2223 crystals were grown by the traveling-solvent floating-zone method,
as described in Ref.~\onlinecite{Giannini-2004}. In order to achieve optimal
doping (OPT), the crystals were annealed during 10 days at 500$^{\circ}$C in
20~bar oxygen partial pressure. This thermal treatment produced a sharpening of
the superconducting transition with respect to the as-grown condition. We
considered the peak position in the temperature derivative of the low-field
susceptibility and in magnetization data as the criteria to determine $T_c$.
Both determinations yielded the same $T_c=(110.5 \pm 0.5)$~K \cite{Piriou-2008}.
The transition width, estimated as the FWHM of the susceptibility and
magnetization peaks, typically ranges between 0.6 and 2~K from sample to sample.
The structural and superconducting properties of OPT crystals of the same batch
as the ones studied here are reported in Refs~\onlinecite{Giannini-2005,
Giannini-2008, Piriou-2008, Piriou-2007}. X-ray diffraction measurements have
revealed the purity and the high crystalline order of the samples
\cite{Giannini-2008}. Resistivity measurements showed a single and sharp
superconducting transition. The sample growth and thermal treatment parameters
were optimized, in order to suppress the intergrowth of the Bi-2212 phase. For
the OPT samples studied here, Bi-2212 intergrowth, if present, represents less
than 1\% of the sample volume \cite{Piriou-2008}.

\begin{figure}[tb]
\includegraphics[width=0.7\columnwidth]{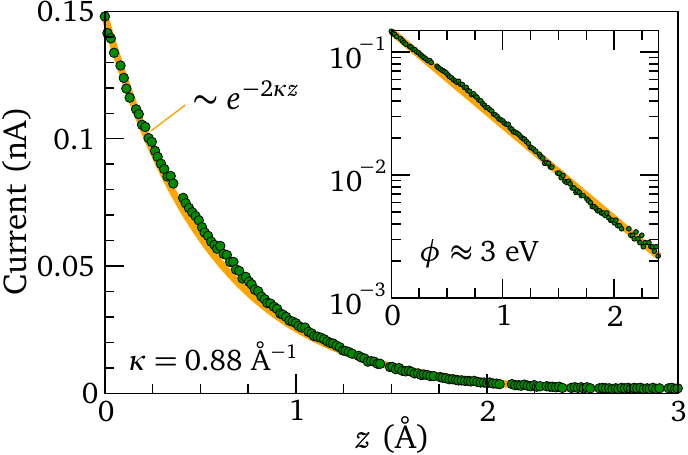}
\caption{\label{fig:fig01}
Decrease of the tunnel current with increasing tip-sample distance $z$ for one
of our junctions (circles). The solid line is an exponential fit. Inset: same
data on a log scale.
}
\end{figure}

For the measurements, we used a home-built STM with ultrahigh-vacuum
environment and $^{3}$He base temperature \cite{Kugler-2000a}.
Electrochemically etched iridium tips served as the ground electrode. The
bias voltage $V$ was applied to the sample, such that negative (positive) bias
refers to occupied (empty) sample states. Differential-conductance spectra were
acquired using a lock-in \cite{Fischer-2007}. The $dI/dV$ measurements were
performed at fixed tip-sample distance, determined by regulation current and
voltage of 0.6~nA and 600~mV, and a lock-in excitation amplitude of 2~mV. The
samples were cleaved at room temperature at $1$--$5 \times 10^{-9}$~mbar
pressure, and cooled down to 2~K in 10 hours time. High-quality tunnel junctions
were obtained in this way, as illustrated in Fig.~\ref{fig:fig01}. In a
high-quality junction, the current depends exponentially on the relative
tip-sample distance, namely $I\propto e^{-2\kappa z}$. The decay constant
$\kappa$ is related to the apparent barrier height $\phi=\hbar^2\kappa^2/(2m)$.
Our tunnel junctions have typically $\phi=3$--$4$~eV, much larger than the
regulation voltage. We started all our runs of measurements with tests like the
one shown in Fig.~\ref{fig:fig01}.

\subsection{Data statistics and systematics of spectral features}

\begin{figure*}[tb]
\includegraphics[width=0.85\textwidth]{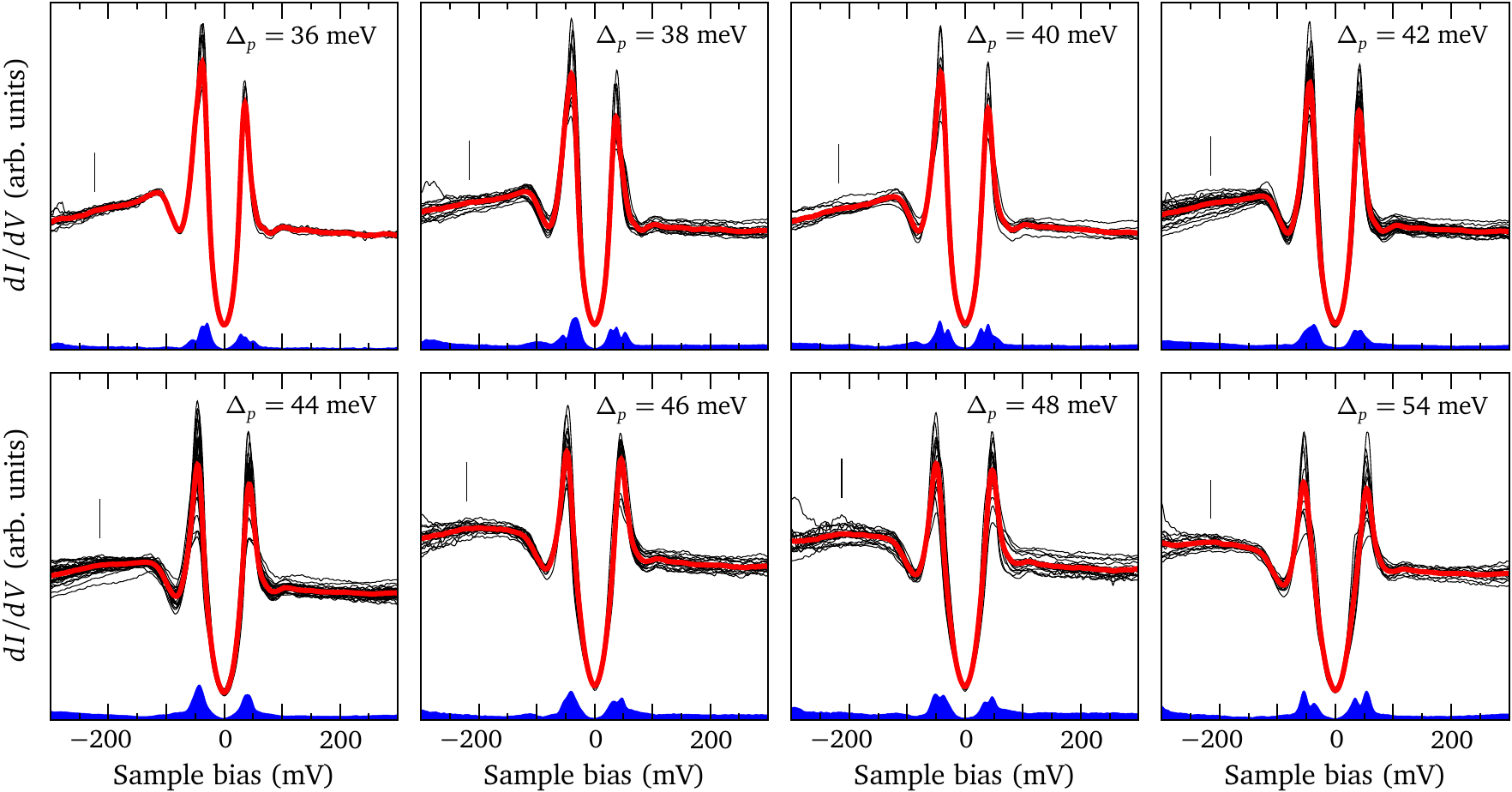}
\caption{\label{fig:fig02}
The complete experimental dataset considered in this study. The raw Bi-2223
tunneling conductance spectra were grouped according to their peak-to-peak gap
$\Delta_p$, and were normalized in order to have the same spectral weight in the
range of the figure (black curves). The red curves are the average spectra, and
the shaded blue curves show the standard deviation of the distribution of
tunneling conductances at each bias. The vertical bars indicate a weak feature,
possibly related to the VHS of the inner-layer band (see Sec.~\ref{sec:theory}).
}
\end{figure*}

As largely documented in the literature, Bi-based cuprate samples, even with a
sharp superconducting transition, present inhomogeneous spectroscopic properties
on the surface \cite{Fischer-2007}. The nanoscale variations of typical spectral
features in Bi-2223 were studied previously by mapping the local $dI/dV$ curves
\cite{Jenkins-2009}. It was found, in particular, that the local gap $\Delta_p$
presents spatial variations in register with the crystalline structure. The
present paper is focused on studying relevant spectral features as a function of
$\Delta_p$. Since the results in several OPT samples are similar, for the
present study we considered 150 spectra measured at different locations of a
given OPT sample. The local $dI/dV$ curves were sorted by half the peak-to-peak
gap $\Delta_p = (\Delta_p^+ - \Delta_p^-)/2$, where $\Delta_p^+$ ($\Delta_p^-$)
is the energy of the coherence peak at positive (negative) bias.
Figure~\ref{fig:fig02} shows the set of spectra, as well as the average spectra
for each $\Delta_p$ (the latter were already published in
Ref.~\onlinecite{Levy-2008}). The standard deviation to the average is also
shown as a function of bias. It is generally $\sim5$\%, except close to the
coherence peaks where it increases up to 20\%. This is a hallmark of the lock-in
technique, which is most reliable for slowly varying $I(V)$ curves.

The prominent spectral features of Bi-2223 are evident from the average $dI/dV$
curves in Fig.~\ref{fig:fig02}. All spectra present a $d$-wave shape at low
bias, very well developed coherence peaks, and a dip at energies larger than
$\Delta_p$. Another remarkable property is a strong electron-hole asymmetry,
characterized by a stronger dip and a greater spectral weight for the occupied
states. This is manifested by higher coherence peaks and an enhanced conductance
background at negative bias. Observed in most high $T_c$'s, this phenomenology
contrasts with the electron-hole symmetric spectra of classical superconductors.

The electron-hole asymmetry evolves monotonically with $\Delta_p$. The
conductance background for occupied states becomes steeper on decreasing the
gap. The trend is particularly evident when considering data in a bias range one
order of magnitude larger than $\Delta_p$. This evolution may be due to a
variation of the correlation effects with the pairing strength
\cite{Anderson-2006}. Since our model does not consider such correlations, we
will restrict our fits to the bias interval $[-150,+150]$~mV. Moreover, the
asymmetry in the height of the coherence peaks systematically increases when
decreasing $\Delta_p$. We discuss the role of the VHS in this phenomenology in
Secs.~\ref{sec:model} and \ref{sec:discussion}. The line shape of the coherence
peaks also follows a monotonic trend. The peaks sharpen and become taller on
approaching the Fermi energy. The evolution of the dip is of particular
importance for this work. This feature is strongly developed in Bi-2223, in
comparison to other Bi-based cuprates, and is noticeably electron-hole
asymmetric. The energy difference between the dip minimum and the coherence peak
maximum decreases on increasing the gap \cite{Jenkins-2009}, as we will discuss
in the following sections.

Finally, the low-energy conductance is similar for all values of $\Delta_p$.
Close to the Fermi level, the spectra are electron-hole symmetric and present a
slightly rounded shape. A V-shaped conductance is characteristic of a $d$-wave
superconductor at zero temperature. Some rounding off and a finite zero-bias
conductance are expected when considering thermal and measurement
broadening---the latter due to the finite amplitude of the lock-in excitation
and to electronic noise---and some residual impurity scattering. Our fitting
procedure takes these factors into account.

\section{Strong-coupling model and fitting procedure}
\label{sec:theory}

\subsection{STM tunneling conductance and LDOS}

The theory of tunneling in superconductors was originally meant for planar
junctions involving classical $s$-wave superconductors, with a structureless
normal-state DOS \cite{Schrieffer-1963}. The theory of Tersoff and Hamann for
the STM \cite{Tersoff-1983}, on the other hand, was not developed for
superconductors. The approach of Ref.~\onlinecite{Tersoff-1983} can be extended
to describe STM measurements in superconductors \cite{Fischer-2007}, and leads
to the paradigm that the differential conductance is a measure of the
thermally broadened electron LDOS:
    \begin{equation}\label{eq:STM}
        \frac{dI(\vec{r})}{dV}=M\int_{-\infty}^{\infty} d\omega d\varepsilon\,
        N(\bm{r},\omega)[-f'(\omega-\varepsilon)]g_{\sigma}(\varepsilon-eV).
    \end{equation}
This expression applies if the current is dominated by single-particle
tunneling. $N(\bm{r},\omega)$ is the sample LDOS at the position $\bm{r}$ of the
STM tip, $f'$ is the derivative of the Fermi function, and $M$ is a
tip-dependent constant. An extra Gaussian broadening by the function
$g_{\sigma}$ takes into account the finite experimental resolution, with
$\sigma$ the half width at half maximum. In addition to electronic noise, the
sources of broadening are the lock-in ac modulation and the averaging of several
similar spectra (see Sec.~\ref{sec:experiment}). The strict proportionality of
$dI/dV$ and $N(\bm{r},eV)$ is recovered in the limits of zero temperature and
$\sigma=0$, where both $-f'$ and $g_{\sigma}$ become delta functions.

If the work function is much larger than the typical energies of interest (in
our case, 3~eV compared with 0.1--0.2~eV; see Fig.~\ref{fig:fig01}), all Bloch
waves decay exponentially outside the sample surface with a similar decay
constant $\kappa$. The LDOS at $\bm{r}$ reduces to
$N(\bm{r},\omega)\propto\exp(-2\kappa z)N(\omega)$, with $N(\omega)$ the bulk
two-dimensional DOS, assumed translation invariant in the $(x,y)$ plane for
simplicity. Consistently, the current must decrease exponentially with $z$, as
confirmed in Fig.~\ref{fig:fig01}. The $z$ dependence is irrelevant in
spectroscopic measurements, and $N(\bm{r},\omega)$ in Eq.~(\ref{eq:STM}) can be
replaced by $N(\omega)$, with a redefinition of the constant $M$. We then
calculate the two-dimensional DOS as the integral of the electron spectral
function:
    \begin{equation}\label{eq:DOS}
        N(\omega)=\frac{2}{N}\sum_{\bm{k}}\left(-1/\pi\right)\text{Im}\,
        G_{11}(\bm{k},\omega).
    \end{equation}
$G_{11}$ is the first component of the Green's function in the Nambu
representation, and $N$ is the number of $\bm{k}$ points in the two-dimensional
Brillouin zone. In a superconductor characterized by a BCS gap
$\Delta_{\bm{k}}$ and inelastic scattering processes, it is convenient to write
$G_{11}$ in terms of the $2\times2$ matrix self-energy
$\hat{\Sigma}(\bm{k},\omega)$, in the form
    \begin{equation}\label{eq:G11}
        G_{11}=\left[\omega-\xi_{\bm{k}}+i\Gamma-\Sigma_{11}
        -\frac{(\Delta_{\bm{k}}+\Sigma_{12})^2}
        {\omega+\xi_{\bm{k}}+i\Gamma-\Sigma_{22}}\right]^{-1}.
    \end{equation}
$\Sigma_{11}(\vec{k},\omega)$ and $\Sigma_{22}(\vec{k},\omega)$ describe the
renormalization and damping of the Bogoliubov quasiparticles in the electron and
hole branches, respectively, while the ``anomalous'' self-energy
$\Sigma_{12}(\vec{k},\omega)$ describes scattering effects in the pairing
channel. The expression of $\hat{\Sigma}$ is provided in the next subsection. If
$\hat{\Sigma}=0$, Eq.~(\ref{eq:G11}) reduces to the BCS Green's function, with
$\xi_{\bm{k}}\equiv\varepsilon_{\vec{k}}-\mu$ the noninteracting electron
dispersion, $\mu$ the chemical potential, and $\Gamma$ a phenomenological
scattering rate \cite{Dynes-1978}. We use a tight-binding model for the band
$\varepsilon_{\vec{k}}$, which reads (setting the lattice parameter $a\equiv 1$)
    \begin{multline}\label{eq:xik}
        \varepsilon_{\bm{k}}\equiv\sum_{\bm{r}}t(|\bm{r}|)\,e^{i\bm{k}\cdot\bm{r}}
        =2t_1(\cos k_x+\cos k_y)+\\4t_2\cos k_x\cos k_y+2t_3(\cos 2k_x+\cos 2k_y)+\\
        4t_4(\cos 2k_x\cos k_y+\cos k_x\cos 2k_y)+\\4t_5\cos 2k_x\cos 2k_y.
    \end{multline}
Note that our conventions for the signs and magnitudes of the hopping amplitudes
$t_i$ differ from those used in Ref.~\onlinecite{Eschrig-2000}. The $d$-wave gap
is $\Delta_{\bm{k}}=\Delta_0(\cos k_x-\cos k_y)/2$.

Bi-2223 has three CuO$_2$ layers per unit cell, hence three bands at the Fermi
level \cite{Mori-2002}. Recent photoemission studies suggest that the bands form
a nearly degenerate doublet, attributed to the outer CuO$_2$ layers, and a
single band associated with the inner CuO$_2$ layer \cite{Ideta-2010a,
*Ideta-2010b, Xu-2010}. The inner-layer band is seen $\sim70$~meV
($\sim200$~meV) below the doublet in the nodal (antinodal) direction. We found
that a one-band model focusing on the doublet, with fewer adjustable parameters
than a three-band model, is sufficient to fit the spectra in the range
$|\omega|<150$~meV. This can be understood, since the doublet carries $2/3$ of
the spectral weight, and the VHS of the inner-layer band lies $\sim200$~meV
below that of the doublet, at the border of our measurement window. We expect
that the modifications induced in the theoretical spectrum by using a multi-band
description would be marginal at low energies. That said, we note that the
average spectra in Fig.~\ref{fig:fig02} systematically present a weak structure
at negative bias, between $-215$ and $-225$~mV, which might be the signature of
the inner-layer VHS. Although a definitive assessment is not possible at this
stage, this observation confirms that possible multiband effects are likely to
be small.

\subsection{Bogoliubov quasiparticles coupled to spin fluctuations}

The theoretical investigation of the coupling between Bogoliubov quasiparticles
and spin fluctuations began with the study of the superfluid transition in
$^3$He (see Ref.~\onlinecite{Tewordt-1973} and references therein), and was
revived after the discovery of high-$T_c$ superconductors \cite{Scalapino-1986,
Lenck-1990, Monthoux-1991}. The minimal model to describe the effects of this
coupling is
    \begin{multline}\label{eq:Sigma}
        \hat{\Sigma}(\bm{k},\omega)=-\frac{1}{N}\sum_{\bm{q}}
        \frac{1}{\beta}\sum_{i\Omega_n}
        g^2\chi_s(\bm{q},i\Omega_n)\\ \times\left.
        \hat{\mathscr{G}}_0(\bm{k}-\bm{q},i\omega_n-i\Omega_n)
        \right|_{i\omega_n\to\omega+i0^+}.
    \end{multline}
$\hat{\mathscr{G}}_0$ is the $2\times2$ Nambu-BCS-Matsubara Green's function in
the absence of coupling, $\chi_s$ is the spin susceptibility, $i\omega_n$ and
$i\Omega_n$ are the fermionic and bosonic Matsubara frequencies, respectively,
$\beta=(k_{\text{B}}T)^{-1}$ is the inverse temperature, and $g$ is the coupling
parameter. Equation~(\ref{eq:Sigma}) can be obtained from perturbation theory in
the electron-spin coupling \cite{Monthoux-2004}; it can also be viewed as a
simplified, non-self-consistent version of a conserving strong-coupling theory
\cite{Wermbter-1993}.

Following Ref.~\onlinecite{Eschrig-2000}, we use a separable phenomenological
expression for $\chi_s$ in the superconducting state. In the energy range of
interest (below $\sim 150$~meV), we assume that the spin response is dominated
by a resonance at energy $\Omega_s$, near the antiferromagnetic vector
$\bm{Q}=(\pi,\pi)$:
    \begin{equation}\label{eq:chis}
        \chi_s(\bm{q},i\Omega_n)=W_s\,F(\bm{q})\int_{-\infty}^{\infty} d\varepsilon\,
        \frac{I(\varepsilon)}{i\Omega_n-\varepsilon}.
    \end{equation}
We choose the real functions $F(\bm{q})$ and $I(\varepsilon)$ such that
$(1/N)\sum_{\bm{q}}F(\bm{q})= \int_0^{\infty}d\varepsilon\,I(\varepsilon)=1$.
$W_s$ thus stands for the momentum and frequency integrated spectral weight of
the resonance:
	\begin{equation}
		W_s=\frac{1}{N}\sum_{\bm{q}}\int_0^{\infty}d\omega\,
		\left(-1/\pi\right)\,\text{Im}\,\chi_s(\bm{q},\omega).
	\end{equation}
The function $F(\bm{q})$ is Lorentzian-like, peaked at $\bm{q}=\bm{Q}$, with
half width at half maximum $\Delta q$:
    \begin{equation}\label{eq:F}
        F(\bm{q})=\frac{F_0}{\sin^2\left(\frac{q_x-Q_x}{2}\right)
        +\sin^2\left(\frac{q_y-Q_y}{2}\right)+(\Delta q/4)^2}.
    \end{equation}
The constant $F_0$ ensures the normalization of $F(\bm{q})$. In
Ref.~\onlinecite{Eschrig-2000}, the resonance was assumed to be sharp in energy,
so that its energy distribution was
$I(\varepsilon)=\delta(\varepsilon-\Omega_s)-\delta(\varepsilon+\Omega_s)$.
Indeed, neutron scattering measurements suggest that the resonance is resolution
limited in Y-123 \cite{Rossat-Mignod-1991}. We use a slightly more general form,
    \begin{equation}\label{eq:I}
        I(\varepsilon)=I_0\big[L_{\Gamma_s}(\varepsilon-\Omega_s)
        -L_{\Gamma_s}(\varepsilon+\Omega_s)\big],
    \end{equation}
where $L_{\Gamma}(\varepsilon)=(\Gamma/\pi)/(\varepsilon^2+\Gamma^2)$ is a
Lorentzian, and $I_0$ ensures the normalization of $I(\varepsilon)$. The form
(\ref{eq:I}) accounts for a finite lifetime $\tau_s\sim\Gamma_s^{-1}$ of the
spin mode. Neutron scattering experiments indicate that the resonance is
somewhat broader in Bi-2212 (Ref.~\onlinecite{Fong-1999}) and Bi-2223
(Ref.~\onlinecite{Pailhes-2004}) than in Y-123, and would be consistent with
$\Gamma_s\approx 4$--8~meV. Alternatively, Eq.~(\ref{eq:I}) may be regarded as a
way to incorporate the observed dispersion of the resonance,\cite{Bourges-2000,
*Reznik-2004} which broadens the mode into a band of width $\Gamma_s$.

It is convenient to write the Matsubara Green's function in Eq.~(\ref{eq:Sigma})
using the spectral representation,
    \begin{equation}\label{eq:G0}
        \hat{\mathscr{G}}_0(\bm{k},i\omega_n)=\int_{-\infty}^{\infty} d\varepsilon\,
        \frac{\hat{A}(\bm{k},\varepsilon)}{i\omega_n-\varepsilon}.
    \end{equation}
Taking into account the phenomenological scattering rate $\Gamma$ appearing in
Eq.~(\ref{eq:G11}), the spectral function can be expressed in terms of
Lorentzian functions,
    \begin{equation}\label{eq:A}
        \hat{A}(\bm{k},\varepsilon)=
        \hat{u}_{\bm{k}}L_{\Gamma}(\varepsilon-E_{\bm{k}})
        +\hat{v}_{\bm{k}}L_{\Gamma}(\varepsilon+E_{\bm{k}}),
    \end{equation}
where
    \begin{equation}\label{eq:uvE}
        \hat{u}_{\bm{k}}=\frac{1}{2}\begin{pmatrix}
        1+\frac{\xi_{\bm{k}}}{E_{\bm{k}}}  & \frac{\Delta_{\bm{k}}}{E_{\bm{k}}}\\
        \frac{\Delta_{\bm{k}}}{E_{\bm{k}}} & 1-\frac{\xi_{\bm{k}}}{E_{\bm{k}}}
        \end{pmatrix},\quad \hat{v}_{\bm{k}}=\openone-\hat{u}_{\bm{k}},
    \end{equation}
and $E_{\bm{k}}=\sqrt{\xi_{\bm{k}}^2+\Delta_{\bm{k}}^2}$. After inserting
Eqs.~(\ref{eq:chis}) and (\ref{eq:G0}) in Eq.~(\ref{eq:Sigma}), one performs the
sum over Matsubara frequencies with the standard technique \cite{Mahan-2000},
and using Eqs.~(\ref{eq:I}) and (\ref{eq:A}), one obtains the self-energy on the
real-frequency axis:
    \begin{multline}\label{eq:Sigma1}
        \hat{\Sigma}(\bm{k},\omega)=\frac{\alpha^2}{N}\sum_{\bm{q}}F(\bm{q})
        [\hat{u}_{\bm{k}-\bm{q}}B(\omega,E_{\bm{k}-\bm{q}})\\
        +\hat{v}_{\bm{k}-\bm{q}}B(\omega,-E_{\bm{k}-\bm{q}})].
    \end{multline}
We have introduced the dimensionless coupling constant
    $
        \alpha^2\equiv(g/\Lambda)^2W_sI_0
    $,
with $\Lambda$ a characteristic energy scale of the model, which we take as the
nearest-neighbor hopping $t_1$. For definiteness, the function $B(\omega,E)$ is
derived in the Appendix. The self-energy (\ref{eq:Sigma1}) is a convolution in
momentum space, and can therefore be efficiently evaluated numerically on dense
$\vec{k}$-point meshes, using fast Fourier transforms.

The use of a coupling constant $\alpha$ comprising the spectral weight of the
resonance, instead of the coupling $g$, is more convenient for our purposes. The
strength of the self-energy effects---hence the strength of the dip in the
tunneling spectrum---is governed by the product $g^2W_s$, and $W_s$ strongly
depends on the momentum width $\Delta q$ of the resonance. In the original
formulation \cite{Eschrig-2000}, both $g$ and $\Delta q$ strongly affect the
dip, whereas here, the dip is controlled mostly by $\alpha$, and depends weakly
on $\Delta q$. As the model parameters will be determined by least-squares fits,
we expect a simpler landscape by avoiding that different parameters have the
same influence on the theoretical spectrum.

\subsection{Discussion of the model}
\label{sec:model}

The model has fifteen parameters, including the multiplicative constant $M$ in
Eq.~(\ref{eq:STM}), thirteen of which are determined by least-squares fitting.
The two fixed parameters are the temperature (set to the experimental value
$T=2$~K) and the Gaussian broadening $\sigma$ (set to 4~meV). We constrain the
scattering rate $\Gamma$ to be larger than 1~meV for the numerical stability of
the momentum sum in Eq.~(\ref{eq:DOS}). The scale of the $d$-wave gap is set by
$\Delta_0$ and modified by the coupling to the spin resonance, as discussed
further below.

\begin{figure}[tb]
\includegraphics[width=0.9\columnwidth]{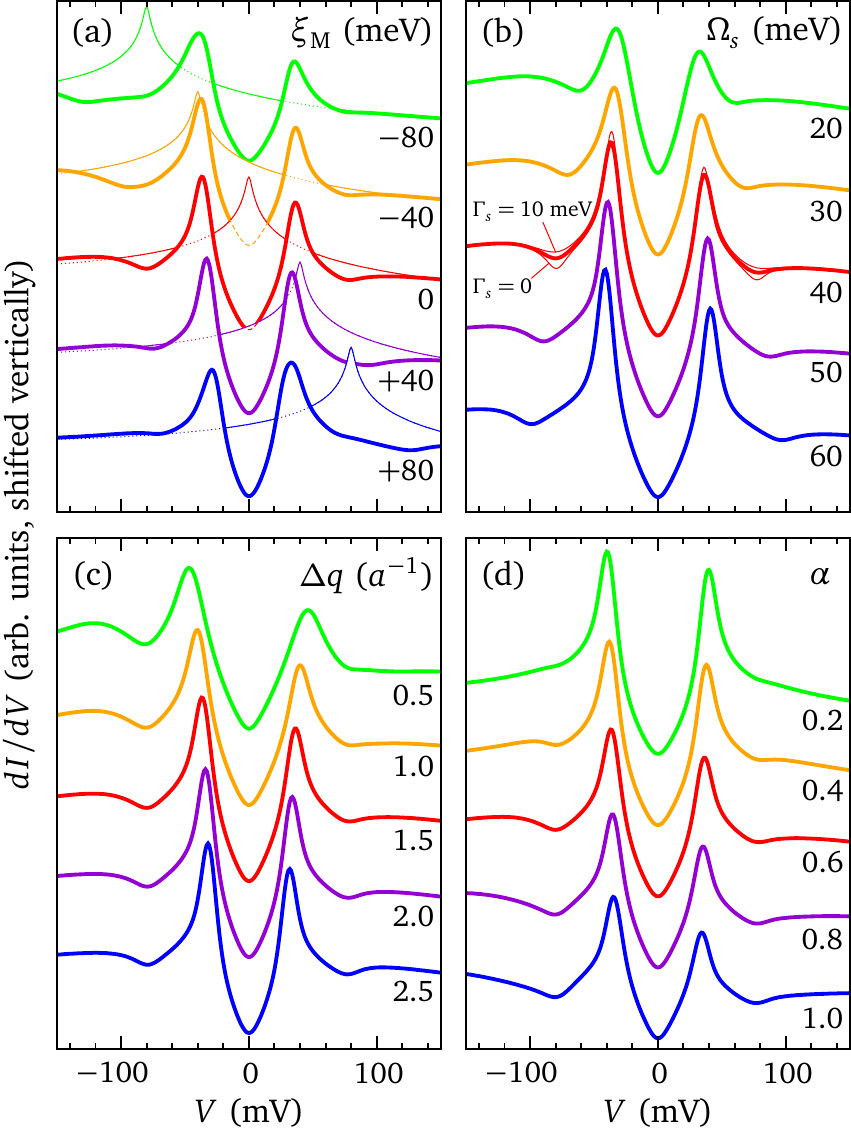}
\caption{\label{fig:fig03}
Evolution of the theoretical tunneling spectrum with varying five model
parameters. The base parameters are (all energies in meV) $M=1$, $\Gamma=2$,
$t_{1\!-\!5}=(-250,80,0,0,0)$, $\mu=-320$, $\Delta_0=40$, $\Omega_s=40$,
$\Gamma_s=5$, $\Delta_q=1.5/a$, $\alpha=0.6$, and they correspond to the middle
red curve in each series. The $\vec{k}$-point mesh contains $1024\times1024$
points. In (a), the chemical potential is varied to move the Van Hove
singularity from $-80$ to $+80$~meV; the thin lines show the corresponding
noninteracting DOS (i.e., with $\Delta_0=0$ and $\alpha=0$). (b), (c), and (d)
show the result of varying the resonance energy $\Omega_s$, $\vec{q}$-space
width $\Delta q$, and the coupling $\alpha$, respectively. The effect of
changing $\Gamma_s$ is illustrated by the thin curves in (b).
}
\end{figure}

The band parameters $t_{1\!-\!5}$ and the chemical potential $\mu$ determine the
noninteracting DOS, the band filling, and the Fermi surface. The tunneling
spectrum is very sensitive to the energy of the VHS given by the dispersion at
the M point $(\pi,0)$, $\xi_{\text{M}}=4(-t_2+t_3+t_5)-\mu$, as illustrated in
Fig.~\ref{fig:fig03}(a). Four trends can be observed: the energy of the VHS
affects (1) the difference in height of the two coherence peaks, (2) the
electron-hole asymmetry of the dip, (3) the overall height of the coherence
peaks, and (4) the half peak-to-peak gap $\Delta_p$. (1) is due to the
superconducting gap, pushing the VHS farther down if initially at negative
energy, and farther up in the opposite case. The coherence peak closest to the
VHS thus carries more spectral weight. One notices that some electron-hole
asymmetry remains when $\xi_{\text{M}}=0$, because a finite $t_2$ breaks the
electron-hole symmetry of the noninteracting DOS. (2) is a consequence of the
dip being reinforced by the VHS and therefore strongest at negative energy if
$\xi_{\text{M}}<0$, and vice versa \cite{Levy-2008}. (3) reveals the
contribution of the VHS to the weight of the coherence peaks, tallest when
$\xi_{\text{M}}=0$. Lastly, (4) is a consequence of $d$-wave symmetry, implying
that the maximum gap on the Fermi surface decreases as the distance between the
Fermi crossing in the antinodal direction and the M point increases.

Varying the resonance energy $\Omega_s$ has three main effects, as shown in
Fig.~\ref{fig:fig03}(b): (1) the dip minimum moves with respect to the closest
coherence peak, (2) the height of the coherence peaks changes, and (3) the gap
$\Delta_p$ varies. (1) indicates that the scattering is strongest near the
energy $\Delta_0+\Omega_s$: Bogoliubov quasiparticles at this energy can easily
decay by emitting a $(\pi,\pi)$ mode, because there are many final states
available near the gap-edge energy $\Delta_0$. One can figure out the typical
energy dependence of the scattering rate by using the limit (\ref{eq:Bsimple})
for the function $B$, and averaging Eq.~(\ref{eq:Sigma1}) over the Brillouin
zone. One arrives at
    \begin{multline}\label{eq:simple_damping_rate}
        -\text{Im}\,\bar{\Sigma}_{11}(\omega)=\frac{\pi}{2}(\alpha t_1)^2
        \big[\theta(\omega-\Omega_s)N_0(\omega-\Omega_s)\\
        +\theta(-\omega-\Omega_s)N_0(\omega+\Omega_s)\big],
    \end{multline}
where $N_0(\omega)$ is the $d$-wave BCS DOS. Since $N_0(\omega)$ peaks at
$\omega\approx\pm\Delta_0$, $-\text{Im}\,\bar{\Sigma}_{11}(\omega)$ peaks at
$\omega\approx\pm(\Delta_0+\Omega_s)$. Due to the confinement of the resonance
around $(\pi,\pi)$, the amplitude of the scattering rate has a marked momentum
dependence \cite{Eschrig-2000, Berthod-2010}. Still,
Eq.~(\ref{eq:simple_damping_rate}) correctly captures the qualitative energy
dependence at all momenta. The effect (2)---increase of the peaks' height with
increasing $\Omega_s$---can also be understood based on
Eq.~(\ref{eq:simple_damping_rate}): the scattering rate is zero for
$|\omega|<\Omega_s$, since Bogoliubov quasiparticles cannot decay, and thus the
coherence peaks are not broadened if $\Omega_s>\Delta_p$. The broadening is
strongest when the peak in $-\text{Im}\,\bar{\Sigma}_{11}$ coincides with
$\Delta_p$, i.e., when $\Omega_s\to0$. The origin of trend (3) is in the
renormalization of the quasiparticle energies, encoded in the real part of the
self-energy. The latter is linear at low energy,
$\text{Re}\,\bar{\Sigma}_{11}(\omega)\approx-\bar{\lambda}\omega$, and the
energy levels are renormalized by a factor $1/(1+\bar{\lambda})$. Performing the
Brillouin-zone average as above leads to
    \begin{equation}\label{eq:lambdabar}
        \bar{\lambda}=(\alpha t_1)^2\frac{1}{2}\int_{-\infty}^{\infty}d\varepsilon\,
        \frac{N_0(\varepsilon)}{(|\varepsilon|+\Omega_s)^2}.
    \end{equation}
If the Fermi point in the antinodal direction is close to the M point, the
scale $\Delta_p$ is given to a good approximation by
$\Delta_0(1+\Psi_{\text{M}})/(1+\lambda_{\text{M}})$. $\lambda_{\text{M}}$ is
the renormalization at the M point, which is typically 30\% larger than the
Brillouin-zone average $\bar{\lambda}$, and
$\Psi_{\text{M}}=\text{Re}\,\Sigma_{12}(\text{M},0)/\Delta_0$ gives the
contribution of the spin resonance to pairing. $\Psi_{\text{M}}$ is a decreasing
function of $\Omega_s$. This can be seen by estimating the Brillouin-zone
average of $\text{Re}\,\Sigma_{12}(\vec{k},0)/\Delta_{\vec{k}}$, which gives
    \begin{equation}\label{eq:Psibar}
        \bar{\Psi}\approx(\alpha t_1)^2\int_0^{\infty}d\varepsilon\,
        \frac{N_0(\varepsilon)}{\varepsilon(\varepsilon+\Omega_s)}.
    \end{equation}
Equation~(\ref{eq:Psibar}) is accurate in the limit $\Delta q\to0$ and for an
electron-hole symmetric band. Looking at the $\Omega_s$ dependence of
$\bar{\Psi}$, one expects a decrease of $\Delta_p$ with increasing $\Omega_s$.
However, this is overcompensated by the faster decrease of $\bar{\lambda}$ with
$\Omega_s$ [see Eq.~(\ref{eq:lambdabar})], and the net result is a slight
increase of $\Delta_p$ with increasing $\Omega_s$, as seen in
Fig.~\ref{fig:fig03}(b). This figure also illustrates the effect of increasing
$\Gamma_s$, namely, a broadening mostly confined to the neighborhood of the dip
minimum.

Figure~\ref{fig:fig03}(c) shows that the momentum width of the resonance affects
the gap renormalization. If $\Delta q\lesssim a^{-1}$, $\Psi_{\text{M}}$ wins
over $\lambda_{\text{M}}$, and $\Delta_p>\Delta_0$, while the opposite happens
if $\Delta q\gtrsim a^{-1}$. The precise value of $\Delta q$ where this change
of behavior takes place depends on the other model parameters. For $\Delta
q\to0$, the renormalizations at M can be evaluated as
    \begin{equation}\label{eq:Dq0}
        \lambda_{\text{M}}=\frac{(\alpha t_1)^2}{(E_{\text{M}}+\Omega_s)^2},\quad
        \Psi_{\text{M}}=\frac{(\alpha t_1)^2}{E_{\text{M}}(E_{\text{M}}+\Omega_s)}
        \quad(\Delta q=0),
    \end{equation}
and indeed $\Psi_{\text{M}}>\lambda_{\text{M}}$ in this case
($E_{\text{M}}=\sqrt{\xi_{\text{M}}^2+\Delta_0^2}$). In the opposite limit,
$\Delta q\to\infty$ or $F(\vec{q})\equiv 1$, the self-energy becomes momentum
independent. We then simply get
    \begin{equation}\label{eq:Dqinf}
        \lambda_{\text{M}}=\bar{\lambda},\quad\Psi_{\text{M}}=0\qquad(\Delta q=\infty),
    \end{equation}
and the gap is reduced by a factor $1/(1+\bar{\lambda})$. The vanishing of
$\Psi_{\text{M}}$ is due to the $d$-wave symmetry of $\Delta_{\vec{k}}$. We
emphasize that the variation in Fig.~\ref{fig:fig03}(c) is performed at
\emph{fixed spectral weight} of the resonance, meaning that the bare coupling
$g$ decreases as $\Delta q$ increases, and $\alpha$ remains unchanged. The value
of $\Delta q$ also influences the height of the coherence peaks: they are
strongly broadened when the renormalized gap is larger than $\Omega_s$, as in
Fig.~\ref{fig:fig03}(b).

\begin{figure}[tb]
\includegraphics[width=0.9\columnwidth]{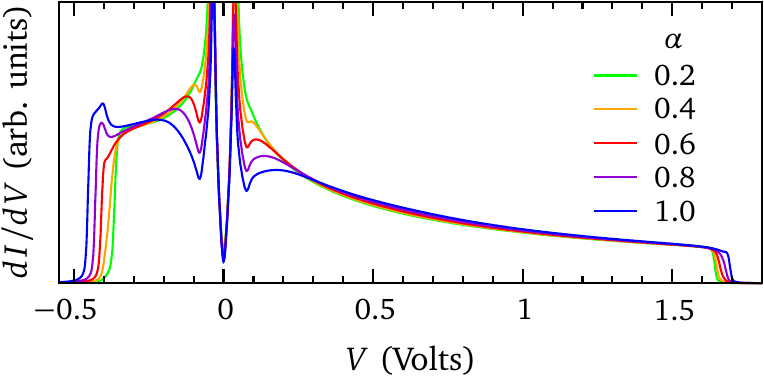}
\caption{\label{fig:fig04}
Same data as Fig.~\ref{fig:fig03}(d) on an expanded energy range covering the
whole bandwidth. Part of the low-energy spectral weight is transferred to the
band edges.
}
\end{figure}

Finally, changing the dimensionless coupling $\alpha$ produces three effects, as
seen in Fig.~\ref{fig:fig03}(d). Increasing $\alpha$ (1) digs the dip, (2)
reduces the gap, and (3) lowers the coherence peaks without broadening them. The
fact that the scattering rate is proportional to $\alpha^2$
[Eq.~(\ref{eq:simple_damping_rate})] explains (1). (2) is due to the fact that
$\Delta q>a^{-1}$ ($\Psi_{\text{M}}<\lambda_{\text{M}}$) in
Fig.~\ref{fig:fig03}(d), as discussed previously; since both $\Psi_{\text{M}}$
and $\lambda_{\text{M}}$ are $\propto\alpha^2$,
$(1+\Psi_{\text{M}})/(1+\lambda_{\text{M}})$ decreases with increasing $\alpha$.
Finally, the effect (3) reflects the removal of low-energy spectral weight by
the coupling to the resonance. This weight is transferred to the ``hump'', but
also over larger energy scales, to the band edges. This is demonstrated in
Fig.~\ref{fig:fig04}, showing the data of Fig.~\ref{fig:fig03}(d) on an expanded
energy range. We further note that part of the spectral weight removed in the
dip is also pushed to \emph{lower} binding energies and, under certain
conditions, can lead to shoulders on the sides of the coherence peaks.

We close this section with a few general remarks. Unlike in the conventional
strong-coupling theory, the model used here does not offer simple relationships
between the parameters and prominent features in the tunneling spectrum. Each
feature is controlled by several parameters. For instance, the gap $\Delta_p$
depends on $\Delta_0$, $\xi_{\text{M}}$, $\Omega_s$, $\Delta q$, and $\alpha$,
as illustrated in Fig.~\ref{fig:fig03}. A direct readout of the parameters by
inspection of the spectra is possible qualitatively, but fitting is required for
extracting accurate values. One peculiarity of the spectrum, however, can be
attributed to a single cause: the electron-hole asymmetry---of the coherence
peaks' height, dip strength, and conductance background---has only one source:
the electron-hole asymmetry of the noninteracting DOS. Therefore, the
qualitative inference of an asymmetric normal DOS \emph{can} be made by direct
inspection of the spectra. Finally, we emphasize that, although the contribution
of the anomalous self-energy to pairing is always positive, $\Delta_p$ may
eventually turn out to be smaller than $\Delta_0$, due to the normal self-energy
renormalization. Namely, the coupling to the spin resonance contributes
positively to pairing, but can nevertheless reduce the preexisting gap.

\subsection{Fitting procedure and variance of the parameters}
\label{sec:fit}

The 13-parameter landscape is too complex for a brute-force fitting approach.
The reason is that the theoretical spectrum depends on properties of the
noninteracting DOS, such as the position and asymmetry of the VHS, that are not
in one-to-one correspondence with the set of hopping amplitudes
$t_i$.\footnote{It is possible to find different sets of tight-binding
parameters which give hardly distinguishable noninteracting DOS in a limited
energy window. For instance, the widely different sets $(\mu, t_1, t_2, t_3,
t_4, t_5)=(-192, -250, 52, -16, 2, 13)$~meV and $(-12, -158, 4, -6, 6, 0)$~meV
yield identical DOS, up to a multiplicative constant, in the energy range
$|\omega|<150$~meV.} A least-squares fit starting with random values of the
parameters will almost certainly end in a local minimum, where the $t_i$ values
do not satisfy physical requirements such as the order of magnitude of the
bandwidth. The fit must therefore be guided with a pinch of physical intuition,
in order to avoid such minima.

In a first step, we have considered the average spectra of Fig.~\ref{fig:fig02}.
For each of them, we searched a set of parameters which (i) is a minimum of the
least-squares function,\footnote{The usual least-squares function,
$F=\sum_i(y_i-Y_i)^2$, measures the vertical distance between two curves
$y_i(x_i)$ and $Y_i(X_i)$. This is not optimal for curves with nearly vertical
segments, such as steep coherence peaks, because a tiny horizontal mismatch is
strongly penalized. We use a different measure of the distance between two
curves: for each point of the first curve, we calculate the shortest Cartesian
distance to any point of the second curve:
$\tilde{F}=\sum_i\min_k[(x_i-X_k)^2+(y_i-Y_k)^2]$. We minimize this function
with the Levenberg-Marquardt algorithm.} (ii) corresponds to a band with the
properties shared by all Bi-based cuprates---band minimum at $\Gamma$, band
maximum at $(\pi,\pi)$, VHS at M---and (iii) gives a holelike Fermi surface
centered at $(\pi,\pi)$. The actual procedure was to search good parameters for
the spectra with $\Delta_p=36$ and $54$~meV (this required a bit of trial and
error), and then to use interpolations between these parameters as seeds to fit
the average spectra with intermediate gaps. The fits were restricted to the
energy window $|\omega|<150$~meV.

\begin{figure}[b]
\includegraphics[width=\columnwidth]{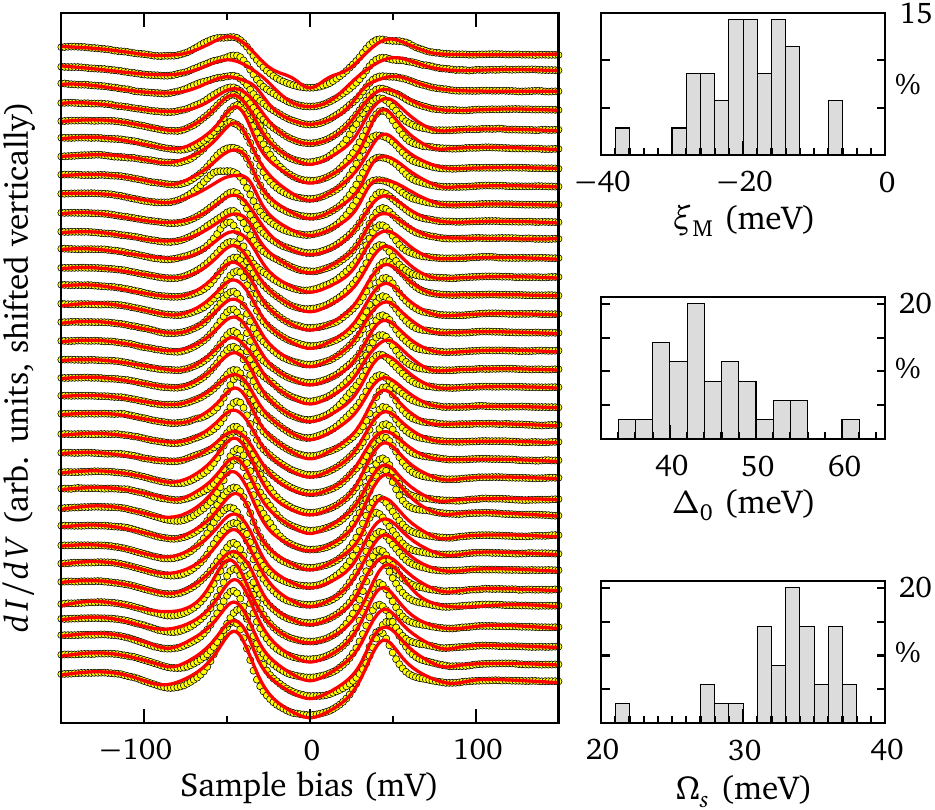}
\caption{\label{fig:fig05}
Series of experimental spectra with a peak-to-peak gap $\Delta_p=44$~meV (yellow
circles), and fits to the strong-coupling model (red lines). The histograms show
the distributions of fitted values for the energy of the VHS, the BCS gap, and
the spin-resonance energy. Similar results are obtained for all series in
Fig.~\ref{fig:fig02}. The standard deviations of the distributions reflect the
sample inhomogeneity, and are given for all parameters in Table~\ref{tab:tab1}.
}
\end{figure}

\begin{figure*}[t]
\includegraphics[width=0.85\textwidth]{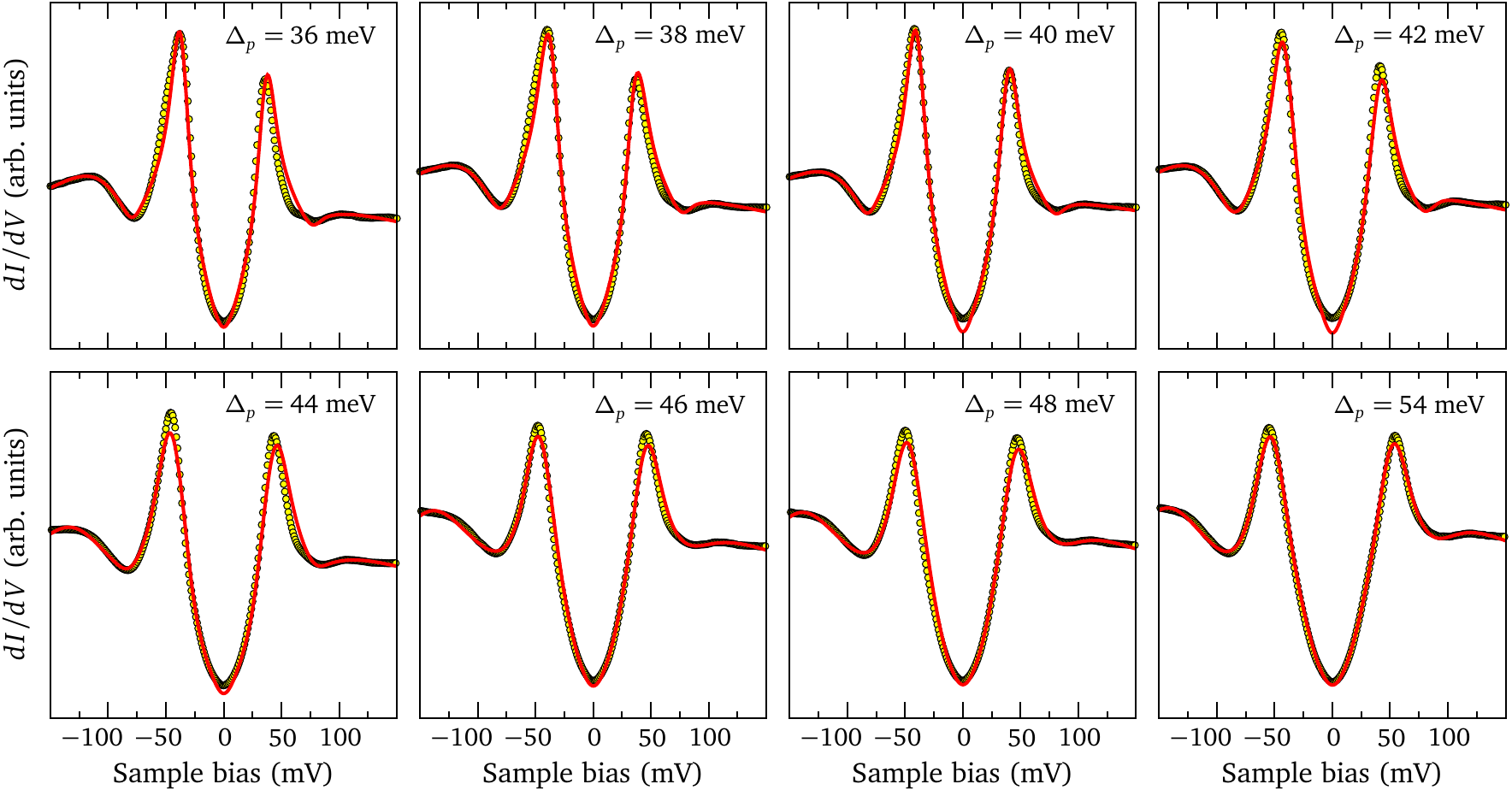}
\caption{\label{fig:fig06}
Average spectra of Fig.~\ref{fig:fig02} (circles) and fits with the model
described in Sec.~\ref{sec:theory} (red curves). The parameters are given in
Table~\ref{tab:tab1}.
}
\vspace{-1.5em}
\end{figure*}

In a second step, we calculated the distributions of the parameters associated
with the sample inhomogeneity, and leading to the fluctuations of individual
spectra around the average (see Fig.~\ref{fig:fig02}). We fitted all individual
spectra, using as seeds the values obtained for the corresponding average
spectrum, and leaving all parameters free to vary without constraint.
Figure~\ref{fig:fig05} shows all individual fits for one of the series in
Fig.~\ref{fig:fig02}, and the distributions of fitted values for three important
parameters. For all parameters, we find that the average of the distribution
coincides with the value obtained by fitting the average spectrum, within the
standard deviation (see Fig.~\ref{fig:fig08} below). This justifies the use of
average spectra to cope with the fluctuations seen in Figs.~\ref{fig:fig02} and
\ref{fig:fig05}.

\section{Results}
\label{sec:results}

\begin{table*}[tb]
\caption{\label{tab:tab1}
Model parameters: for each experimental gap $\Delta_p$ and each parameter, the
first column gives the value obtained by fitting the average spectrum
(Fig.~\ref{fig:fig06}); the number in parentheses is the standard deviation of
the values obtained by fitting all individual spectra, as illustrated in
Fig.~\ref{fig:fig05}.
}
\def\col{@{\extracolsep{\fill}}r@{\extracolsep{1mm}}r}
\begin{tabular*}{0.8\textwidth}{c\col\col\col\col\col\col}
\hline\hline
$\Delta_p$ & \multicolumn{12}{c}{Parameters of the dispersion}\\
(meV) & \multicolumn{2}{c}{$\mu$ (meV)} & \multicolumn{2}{c}{$t_1$ (meV)} &
\multicolumn{2}{c}{$t_2$ (meV)} & \multicolumn{2}{c}{$t_3$ (meV)} &
\multicolumn{2}{c}{$t_4$ (meV)} & \multicolumn{2}{c}{$t_5$ (meV)} \\[0.2em]
\hline
$36$ & $-165$ & $ (56)$ & $-180$ & $ (43)$ & $32$ & $(24)$ & $-11$ & $(10)$ & $  2.2$ & $ (8.6)$ & $-1.6$ & $ (3.7)$ \\
$38$ & $-125$ & $(111)$ & $-161$ & $ (79)$ & $21$ & $(36)$ & $-14$ & $(17)$ & $  5.8$ & $ (9.3)$ & $-1.4$ & $ (6.3)$ \\
$40$ & $-101$ & $(105)$ & $-140$ & $ (85)$ & $ 3$ & $(46)$ & $-23$ & $(21)$ & $ 11.9$ & $(15.5)$ & $-3.4$ & $ (4.1)$ \\
$42$ & $-116$ & $ (51)$ & $-162$ & $ (46)$ & $18$ & $(26)$ & $-17$ & $(13)$ & $  4.3$ & $ (9.2)$ & $-0.8$ & $ (3.5)$ \\
$44$ & $-237$ & $(108)$ & $-206$ & $(217)$ & $56$ & $(50)$ & $-36$ & $(93)$ & $-10.3$ & $(39.9)$ & $27.9$ & $(42.0)$ \\
$46$ & $-306$ & $ (73)$ & $-179$ & $ (66)$ & $59$ & $(43)$ & $-79$ & $(49)$ & $-22.6$ & $(12.6)$ & $56.6$ & $(16.6)$ \\
$48$ & $-340$ & $(108)$ & $-206$ & $(109)$ & $60$ & $(52)$ & $-97$ & $(55)$ & $-26.3$ & $ (9.3)$ & $67.0$ & $(24.9)$ \\
$54$ & $-305$ & $(146)$ & $-256$ & $ (87)$ & $58$ & $(44)$ & $-88$ & $(73)$ & $-35.4$ & $(15.9)$ & $62.9$ & $(25.6)$ \\[0.2em]
\hline\\[-1.2em]
\end{tabular*}
\begin{tabular*}{0.8\textwidth}{c\col\col\col\col\col\col}
 & \multicolumn{12}{c}{Scattering rate, BCS gap, and spin resonance}\\
\phantom{(meV)} & \multicolumn{2}{c}{$\Gamma$ (meV)} & \multicolumn{2}{c}{$\Delta_0$ (meV)} &
\multicolumn{2}{c}{$\Omega_s$ (meV)} & \multicolumn{2}{c}{$\Gamma_s$ (meV)} &
\multicolumn{2}{c}{$\Delta q$ ($a^{-1}$)} & \multicolumn{2}{c}{$\alpha$} \\[0.2em]
\hline
$36$ & $2.0$ & $(0.7)$ & $42.7$ & $(1.0)$ & $36.8$ & $(0.9)$ & $ 1.6$ & $(1.1)$ & $1.39$ & $(0.11)$ & $0.73$ & $(0.14)$ \\
$38$ & $2.1$ & $(0.6)$ & $45.0$ & $(3.7)$ & $34.5$ & $(2.5)$ & $ 0.5$ & $(1.1)$ & $1.46$ & $(0.34)$ & $0.79$ & $(0.19)$ \\
$40$ & $1.0$ & $(0.9)$ & $47.9$ & $(3.4)$ & $34.4$ & $(2.3)$ & $ 1.7$ & $(2.1)$ & $1.54$ & $(0.38)$ & $0.97$ & $(0.29)$ \\
$42$ & $1.0$ & $(0.9)$ & $52.5$ & $(4.9)$ & $29.9$ & $(3.2)$ & $ 1.7$ & $(2.1)$ & $1.65$ & $(0.49)$ & $0.81$ & $(0.18)$ \\
$44$ & $3.6$ & $(2.4)$ & $48.6$ & $(7.3)$ & $33.7$ & $(4.5)$ & $ 4.2$ & $(5.4)$ & $1.15$ & $(0.60)$ & $0.70$ & $(0.27)$ \\
$46$ & $6.6$ & $(1.7)$ & $59.7$ & $(6.4)$ & $21.6$ & $(9.9)$ & $18.6$ & $(7.0)$ & $1.68$ & $(0.69)$ & $1.16$ & $(0.32)$ \\
$48$ & $7.6$ & $(1.8)$ & $60.4$ & $(5.9)$ & $23.1$ & $(9.1)$ & $15.1$ & $(7.2)$ & $1.80$ & $(0.62)$ & $0.97$ & $(0.38)$ \\
$54$ & $8.5$ & $(1.7)$ & $70.7$ & $(6.3)$ & $19.2$ & $(6.8)$ & $13.7$ & $(8.0)$ & $2.17$ & $(0.58)$ & $0.79$ & $(0.35)$ \\[0.2em]
\hline\hline
\end{tabular*}
\vspace{-1em}
\end{table*}

Figure~\ref{fig:fig06} presents the results of our fits to the average spectra,
and Table~\ref{tab:tab1} lists the set of fitted parameters, as well as the
standard deviations. The model can reproduce the relative values of the
conductance at zero bias, on the coherence peaks, and on the background, as they
vary with increasing $\Delta_p$. It also captures the electron-hole asymmetry of
the coherence peaks, and the decrease of this asymmetry with increasing
$\Delta_p$. The electron-hole asymmetry of the dip can be followed as a function
of $\Delta_p$, even when the coherence peaks have become almost symmetric, at
$\Delta_p=54$~meV. Good fits can also be achieved in the full energy range of
Fig.~\ref{fig:fig02}, at the price of a slight deterioration of the fit at low
energies, especially for the largest gaps. The progressive inadequacy of the
model for increasing energy range can have various causes. The presence in the
experimental spectrum of components not considered in the model is one
possibility, for instance, the contribution of the inner-layer band, or an
enhanced spectral weight of the occupied states due to correlations
\cite{Kugler-2006}. Another possibility would be additional scattering
mechanisms at high energy, in particular by the continuum of spin fluctuations
\cite{Eschrig-2003}.

Before discussing the parameters, we emphasize three assertions which are
supported by the quality of the fits in Fig.~\ref{fig:fig06}. First, STS
measures the local \emph{electron} DOS, including band-structure effects: it is
not possible to subtract or divide out the noninteracting DOS by normalization.
Second, the presence of the Van Hove singularity in the noninteracting DOS is
crucial to reproduce the various asymmetries of the spectra: this is the
\emph{only} source of electron-hole asymmetry in the model. Third, a coupling to
the spin resonance can explain quantitatively the redistribution of spectral
weight around the dip energy. The tight localization of the resonance around
$(\pi,\pi)$, as opposed to optical phonons that span the whole Brillouin zone,
plays a key role in producing the correct line shape of the dip.

Table~\ref{tab:tab1} shows that large uncertainties are associated with the
$t_i$'s. This illustrates the weak sensitivity of the theoretical spectrum to
the band parameters (see Sec.~\ref{sec:fit}). The nearest-neighbor hopping $t_1$
varies between $-140$ and $-256$ meV: these numbers fall within the range of
published values for Bi-based cuprates \cite{Norman-1995, Eschrig-2000,
Kordyuk-2003, Hashimoto-2008}. The hopping $t_3$ (second neighbor along the
Cu-Cu direction) is negative like $t_1$, as expected from symmetry
considerations. Likewise, the diagonal hoppings $t_2$ and $t_5$ are both
positive, or the latter is almost zero (we consider the small negative values of
order 1~meV as insignificant). We do not try to interpret the variations of the
$t_i$ with increasing $\Delta_p$, because these variations are comparable with
the typical uncertainties. In fact, the fitted hopping amplitudes should be
regarded as a parametrization of the low-energy DOS, rather than an accurate
determination of the microscopic Hamiltonian. We will see that two properties of
the dispersion which characterize the low-energy DOS, .e.g., the VHS energy and
the Fermi velocity, show comparatively smaller uncertainties, and a systematic
trend with increasing $\Delta_p$.

The scattering rate $\Gamma$ is small with a small variance, and a tendency to
increase with increasing $\Delta_p$. This reflects a trend in the average
spectra to be broader for larger gaps. At energies below $\Omega_s$, $\Gamma$
provides the only broadening mechanism. This parameter is therefore well
constrained by the line-shape around zero bias. At higher energies and for the
large gaps, additional broadening is provided by $\Gamma_s$. Values of $\sim
15$~meV seem somewhat too large for $\Gamma_s$, when compared with available
values for Bi-2212 and Bi-2223 \cite{Fong-1999, Pailhes-2004}, but we note that
their uncertainty is also large. We find that $\Delta_0$ is larger than
$\Delta_p$, and that the difference increases with increasing $\Delta_p$. As
discussed in Sec.~\ref{sec:model}, this is connected with $\Delta q$ being
larger than $1/a$, so that the pairing induced by the spin resonance is
overcompensated by the downward renormalization of the energy levels. The
fitted $\Omega_s$ are anticorrelated with $\Delta_p$, as discussed further
below, and consistently with previous studies \cite{Levy-2008, Jenkins-2009}.
Lastly, the dimensionless parameter $\alpha$ exhibits no clear trend. However
the product $(\alpha t_1)^2$, which controls the coupling strength
[Eq.~(\ref{eq:simple_damping_rate})], increases steadily with increasing
$\Delta_p$. We discuss the coupling strength further in Sec.~\ref{sec:coupling}.

\section{Discussion}
\label{sec:discussion}

\subsection{Electron-hole asymmetry and Van Hove singularity}

\begin{figure}[b]
\includegraphics[width=0.7\columnwidth]{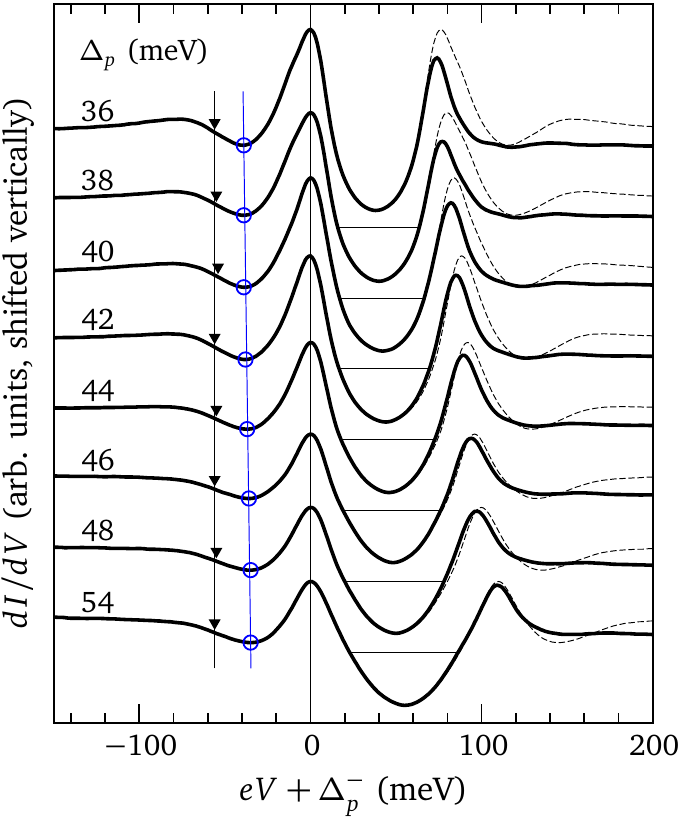}
\caption{\label{fig:fig07}
Average spectra of Fig.~\ref{fig:fig02}, with energies measured from one
coherence peak. The dashed lines show the negative-bias spectrum, mirrored at
positive energies to highlight the electron-hole asymmetry. The circles indicate
the dip minimum, the triangles show the inflection point (peak in $d^2I/dV^2$),
and the straight lines are guides to the eye.
}
\end{figure}

The STS spectra exhibit systematic asymmetries between positive and negative
bias, suggestive of an intrinsic electron-hole asymmetry. The asymmetries
concern the height of the coherence peaks, the strength of the dip, and the
conductance background, as highlighted in Fig.~\ref{fig:fig07}. In the model,
any asymmetry can be traced back to the band structure. For instance, one often
finds that the negative-energy coherence peak is taller if the energy
$\xi_{\text{M}}$ of the VHS is negative [but there are exceptions; see
Fig.~\ref{fig:fig03}(a)]. There is also a simple relationship between
$\xi_{\text{M}}$ and the width and asymmetry of the dip \cite{Levy-2008}, as is
clearly seen in Fig.~\ref{fig:fig03}(a). A wider and/or stronger dip at negative
bias means that $\xi_{\text{M}}<0$. In this respect, the series of spectra in
Fig.~\ref{fig:fig07} present two trends, which seem to have conflicting
implications. On the one hand, the coherence peaks become more symmetric with
increasing $\Delta_p$, suggesting that $\xi_{\text{M}}$ is negative for small
gaps and approaches zero for larger gaps. On the other hand, there is a
tendency for the dip to become wider and more asymmetric with increasing
$\Delta_p$, indicating that the VHS moves to lower energies with increasing
$\Delta_p$.

The fits confirm the latter view, with larger negative values of
$\xi_{\text{M}}$ for larger gaps [Fig.~\ref{fig:fig08}(a)]. This evolution is
consistent with the interpretation that spectra with larger gaps correspond to
more underdoped regions with higher electron densities. Extrapolating our
results, we expect a VHS at positive energy for gap values lower than 24~meV,
i.e., on the strongly overdoped side. We are not aware of any systematic
investigation of the VHS by ARPES in Bi-2223. Reference~\onlinecite{Matsui-2003}
gives one point of comparison, with a dispersion approaching $-25$~meV at
$(\pi,0)$, the value that we obtain for gaps close to $45$~meV. However, this
analysis neglects renormalization effects, and may underestimate
$\xi_{\text{M}}$.

\subsection{Spin-resonance energy}

\begin{figure}[tb]
\includegraphics[width=\columnwidth]{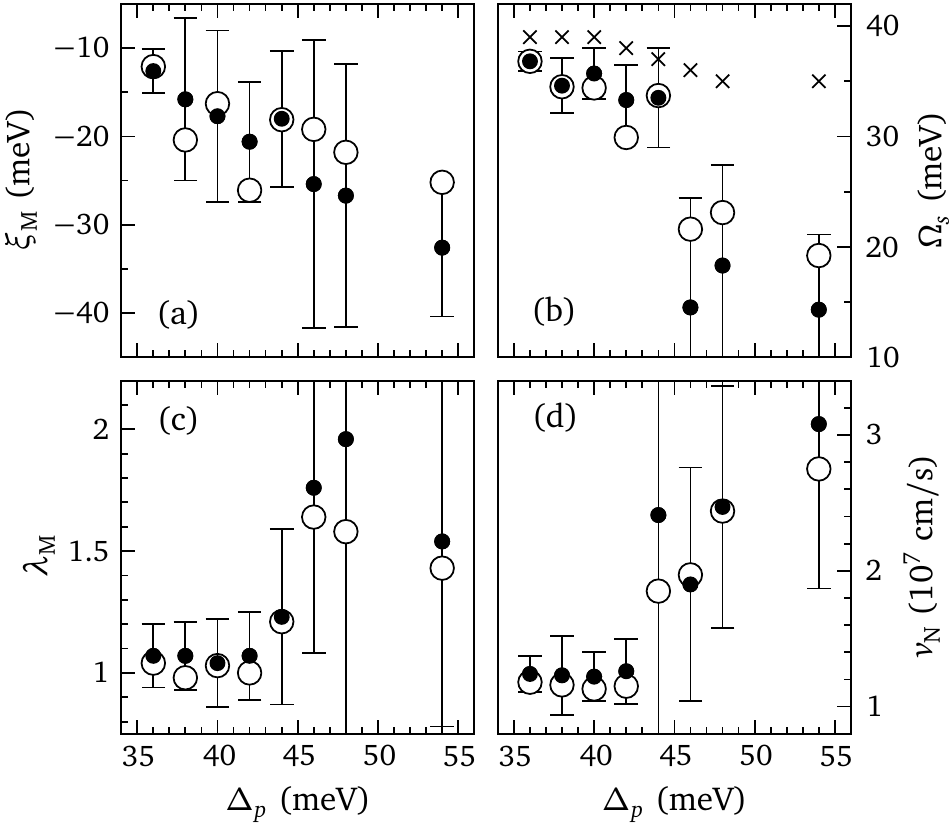}
\caption{\label{fig:fig08}
Evolution of four properties with $\Delta_p$. The empty symbols show the values
obtained by fitting the average spectra. The full symbols with error bars show
the average and standard deviation of the distributions obtained by fitting all
individual spectra. The crosses in (b) represent the peak-to-dip energy
difference of the average spectra, corresponding to the circles in
Fig.~\ref{fig:fig07}. (c) Renormalization factor at the M point,
Eq.~(\ref{eq:lambda_k}). (d) Nodal velocity.
}
\end{figure}

Figure~\ref{fig:fig08}(b) shows the fitted values of $\Omega_s$. For the
smallest gaps, corresponding to spectra with tall and asymmetric coherence
peaks, we find values between 30 and 40~meV, and a good correspondence between
$\Omega_s$ and the peak-to-dip energy difference. Both follow the same
decreasing trend with increasing $\Delta_p$. In classical superconductors, the
phonon energies coincide with peaks in $d^2I/dV^2$ \cite{McMillan-1965}. We
stress that in our spectra, the peak in $d^2I/dV^2$ occurs at an energy
$\sim56$~meV (triangles in Fig.~\ref{fig:fig07}), similar to the values observed
in Bi-2212 \cite{Lee-2006a}. This feature therefore does not provide a good
estimate of $\Omega_s$. This difference with classical superconductors is a
consequence of the $d$-wave symmetry of the gap \cite{Berthod-2010,
Johnston-2010}. For gaps larger than 45~meV, $\Omega_s$ drops abruptly to values
close to 20~meV, in contrast to the peak-to-dip energy, which stays close to
35~meV.

We believe that part of this drop is driven by changes in the spectra going
beyond the scope of the model: the drop is accompanied by other changes, such as
a raise of $\Gamma_s$ and $t_5$ (see Table~\ref{tab:tab1}), suggesting that the
fit has moved to a different region of the parameter space. An obvious change
between the 44 and 46~meV spectra is a falloff of the peak to background ratio.
Increasing $\Gamma$ and/or $\Gamma_s$ cannot account for this, since it would
also reduce the dip to background ratio, which remains unchanged in the spectra.
The compromise is to lower $\Omega_s$, and thus reduce the peak height without
affecting the dip strength [see Fig.~\ref{fig:fig03}(b)], and to tune the
position of the dip minimum by adjusting other parameters [in particular $\Delta
q$; see Fig.~\ref{fig:fig03}(c)]. Another trend is that the coherence peaks
become symmetric for large gaps, and this also drives $\Omega_s$ downward in the
fits.

The range of $\Omega_s$ values in Fig.~\ref{fig:fig08}(b) corresponds well to
the doping evolution observed by neutron scattering in Y-123, where the
spin-resonance energy decreases from $\sim40$~meV at optimal doping to
$\sim25$~meV in underdoped samples \cite{Bourges-1995, Fong-1997}. In Bi-2212,
the resonance is found at 42~meV at optimal doping, and goes down to 34~meV with
strong overdoping \cite{Capogna-2007}, but no data have been reported in the
underdoped region. Without neutron scattering data for Bi-2223, and considering
the large variance of $\Omega_s$ for $\Delta_p>45$~meV, it is difficult to
ascertain whether the decrease of the spin-resonance energy for large gaps is as
sudden as suggested by our fits, or rather more continuous.

\subsection{Coupling strength and renormalization factors}
\label{sec:coupling}

\begin{figure}[tb]
\includegraphics[width=\columnwidth]{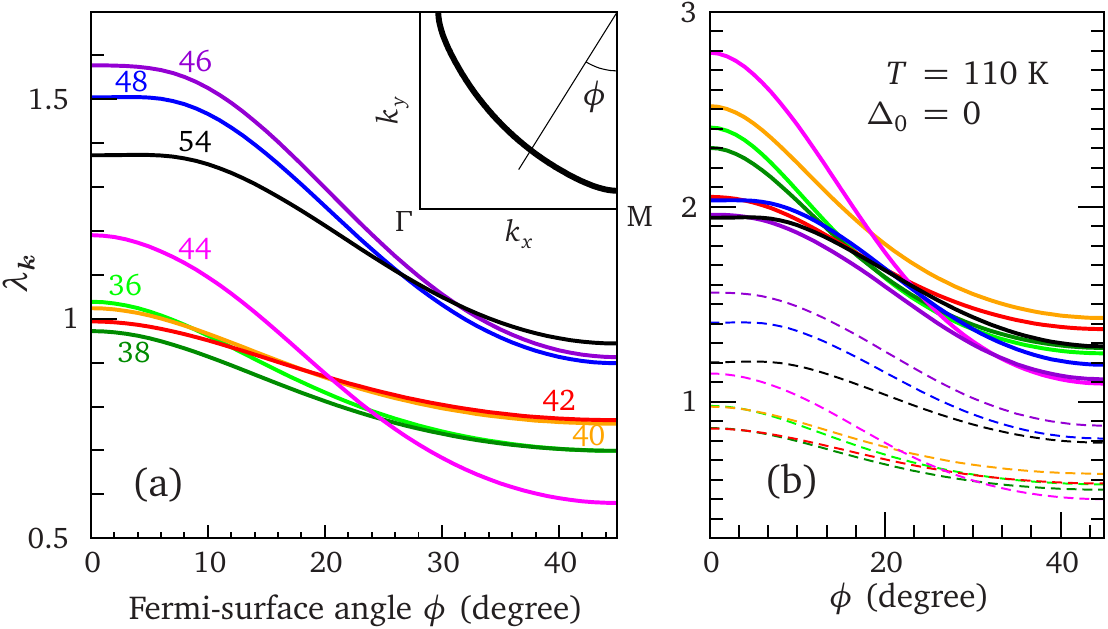}
\caption{\label{fig:fig09}
Renormalization factor $\lambda_{\vec{k}}$ along the renormalized Fermi surface.
(a) Superconducting state: the parameters are those corresponding to the average
spectra of Fig.~\ref{fig:fig06} with the corresponding value of $\Delta_p$
written on each curve. Inset: Fermi surface for $\Delta_p=42$~meV, and
definition of the Fermi-surface angle $\phi$. (b) Normal state: the solid lines
show the value of $\lambda_{\vec{k}}$ obtained by setting $T=110$~K and
$\Delta_0=0$, and keeping the other parameters unchanged. The dashed lines
correspond to $T=110$~K, $\Delta_0=0$, and $\Gamma_s=25$~meV.
}
\end{figure}

The values of the bare coupling strength $g$ obtained from our fits are similar
to those used in Ref.~\onlinecite{Eschrig-2000}. A more meaningful measure of
the strength of self-energy effects is given by the renormalization factor
	\begin{equation}\label{eq:lambda_k}
		\lambda_{\vec{k}}=-\frac{d}{d\omega}\frac{1}{2}\text{Re}\,\left[
		\Sigma_{11}(\vec{k},\omega)+\Sigma_{22}(\vec{k},\omega)\right]_{\omega=0}.
	\end{equation}
Figure~\ref{fig:fig08}(c) shows the evolution of $\lambda_{\vec{k}}$ at the M
point with varying $\Delta_p$, and Fig.~\ref{fig:fig09} shows the anisotropy of
$\lambda_{\vec{k}}$ along the Fermi surface. $\lambda_{\vec{k}}$ is maximal at
the antinodes and minimal at the nodes. The increase of $\lambda_{\text{M}}$
above $\Delta_p=45$~meV is accompanied by an increase of anisotropy: while the
renormalization in antinodal and nodal regions differs by 20--30\% for the
smaller gaps, this increases to 30--50\% for the larger gaps. Note that the
$\lambda_{\vec{k}}$ shown in Fig.~\ref{fig:fig09}(a) are significantly larger,
and more anisotropic, than the values found for Bi-2212 using phonon models
\cite{Johnston-2010a}. This suggests that a fit of phonon models to our STM data
would yield unrealistically large electron-phonon matrix elements.

As the experimental determination of $\lambda$ by ARPES and optical conductivity
are mostly performed in the normal state, we have calculated $\lambda_{\vec{k}}$
at $T=T_c=110$~K. Equation~(\ref{eq:lambdabar}), which approximates the
Brillouin-zone average of $\lambda_{\vec{k}}$, shows that the renormalization
\emph{increases} in the normal state, because the gap in $N_0(\varepsilon)$
closes. Setting $T=110$~K and $\Delta_0=0$, while keeping the other parameters
unchanged, we obtain values of $\lambda_{\vec{k}}$ which are 1.2--2.4 times
larger than in the superconducting state, and more anisotropic
[Fig.~\ref{fig:fig09}(b)]. This calculation overlooks that the transition to the
normal state also affects the spin resonance, which broadens in energy
\cite{Regnault-1995} on warming across $T_c$. This effect can be modeled by
increasing $\Gamma_s$. As an illustration, we show in Fig.~\ref{fig:fig09}(b)
the renormalization calculated with $\Gamma_s=25$~meV (the normal-state value of
$\Gamma_s$ in Bi-2223 has not been determined experimentally). The values of
$\lambda_{\vec{k}}$ are reduced and become similar to the values found in the
superconducting state.

Our renormalization factors compare well with experimental values reported in
the literature. In underdoped and overdoped Bi-2212, a normal-state
renormalization of 1.5 near the antinodal point was determined by ARPES
\cite{Kim-2003}. As the dip feature is stronger in Bi-2223 than in Bi-2212, a
larger value may be expected for the three-layer compound. Indeed, the average
renormalization in the normal state of Bi-2223 at optimal doping was estimated
by fitting a model with a bosonic spectrum to optical data
\cite{vanHeumen-2009a}, and leads to the values 2.18 and 1.75, depending on whether
the full bosonic spectrum or its low-energy part is taken into account,
respectively. In the normal state, but at the nodal point, a renormalization
decreasing from $0.8$ to $0.55$ as a function of increasing hole doping was
measured in Bi-2212 at 120~K \cite{Kordyuk-2006}. Our nodal values for Bi-2223
in the normal state with $\Gamma_s=25$~meV draw a similar trend, decreasing from
$0.8$ to $0.5$ with decreasing gap size. Below $T_c$, nodal values between $0.7$
and $0.9$ are reported for Bi-2212 in Ref.~\onlinecite{Kordyuk-2006}. For
optimally doped Bi-2223, a recent study \cite{Ideta-2013} allows us to estimate a
nodal renormalization of 0.6 at 10~K. These numbers are very close to our
superconducting-state results of Fig.~\ref{fig:fig09}(a). Extracting the
low-temperature antinodal renormalization from ARPES is difficult due to the
gap. A study reported a value of 2 for optimally doped Bi-2212 \cite{Fink-2006},
while our values for Bi-2223 vary between 1 and 1.6.

Figure~\ref{fig:fig08}(d) shows the nodal velocity, calculated using the fitted
parameters and an in-plane lattice constant of 3.825~\AA\ for Bi-2223. The nodal
velocity can be measured directly by ARPES, unlike the renormalization (which
requires an assumption for the bare dispersion). Reference~\onlinecite{Ideta-2013}
reports nodal velocities between 1.5 and 1.7~eV \AA\ for the outer-layer band of
optimally doped Bi-2223, corresponding to $2.3$--$2.6\times10^7$~cm/s. These
values agree well with our results for $\Delta_p\gtrsim44$~meV. In the case of
Bi-2212, values ranging from 1 to $2.5\times10^7$~cm/s as a function of doping
have been reported \cite{Vishik-2010}. The variation seen in
Fig.~\ref{fig:fig08}(d) can therefore be interpreted as reflecting changes in
the local doping level.

\begin{figure}[b]
\includegraphics[width=\columnwidth]{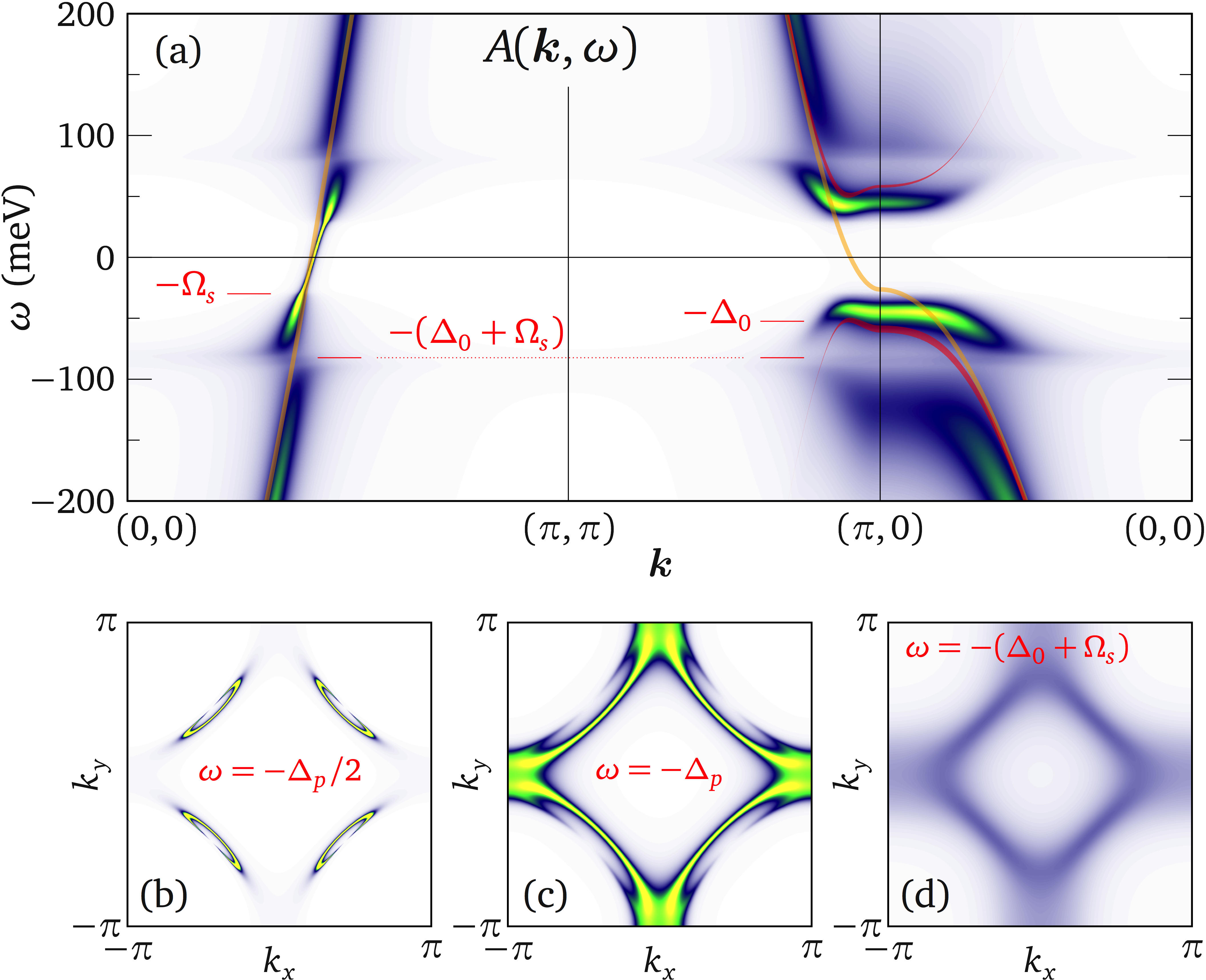}
\caption{\label{fig:fig10}
Spectral function calculated with the parameters fitted to the $\Delta_p=42$~meV
spectrum in Fig.~\ref{fig:fig06}. (a) Along high-symmetry lines in the Brillouin
zone. The color scale shows the variation of the spectral function from zero
(white) to its maximum (yellow). The orange line is the noninteracting
dispersion. The red line shows the BCS dispersion; the width of the line is
proportional to the spectral weight. (b), (c), and (d) Cuts at three
characteristic energies. The color scale is the same in all graphs.
}
\end{figure}

\begin{figure*}[t]
\includegraphics[width=0.8\textwidth]{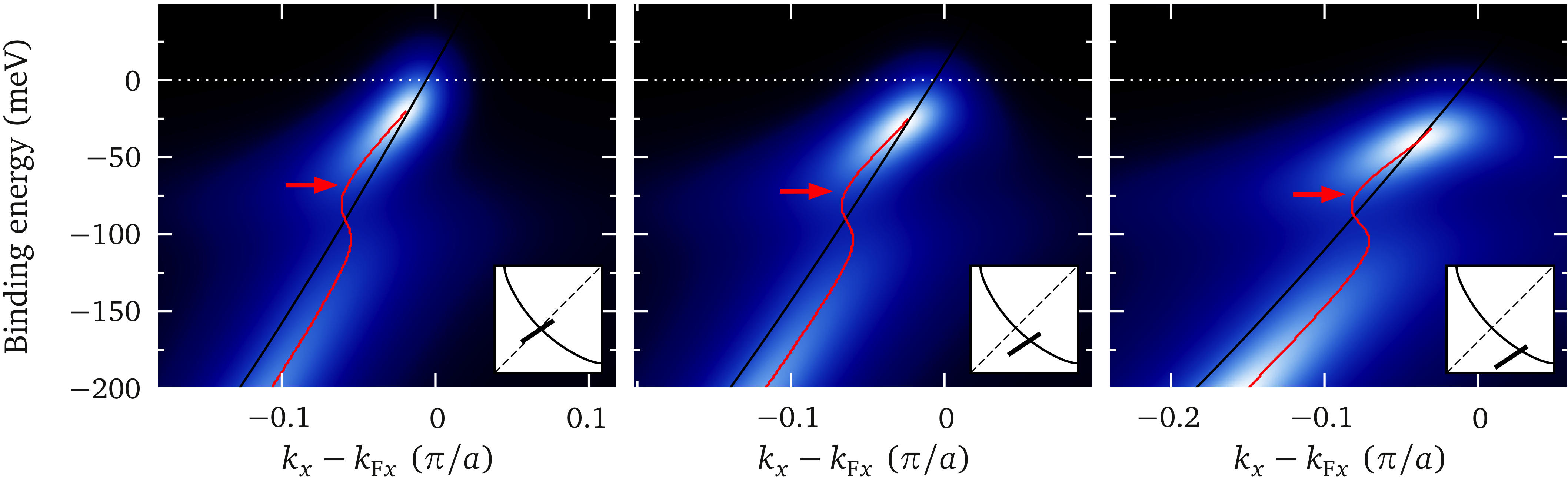}
\caption{\label{fig:fig11}
Simulated ARPES intensity for optimally doped Bi-2223 at $T=10$~K. The spectral
function of Fig.~\ref{fig:fig10} was multiplied by the Fermi function, and
filtered to mimic a momentum resolution of $0.05/a$ and an energy resolution of
18~meV. The color scale and the three momentum cuts correspond approximately to
those of Fig.~2 in Ref.~\onlinecite{Ideta-2008}. The black lines show the
noninteracting dispersion, and the red lines shows the quasiparticle dispersion
(maximum of momentum distribution curves) for energies larger than the gap. The
arrows indicate the ``kink'', where the quasiparticle dispersion deviates from
its low-energy linear behavior.
}
\end{figure*}

\subsection{Spectral function and simulated ARPES intensity}

The parameters determined by fitting STM spectra allow us to calculate the
momentum-resolved spectral function and to make predictions for the measured
ARPES intensity. The generic properties of the spectral function in the model
have been extensively discussed \cite{Eschrig-2000, *Eschrig-2003,
Eschrig-2006}. The main characteristics are summarized in Fig.~\ref{fig:fig10}.
At energies below the onset of scattering, $|\omega|<\Omega_s$, the dispersion
is renormalized downwards, but not broadened. This can be seen most clearly
along the nodal direction $(0,0)$--$(\pi,\pi)$ in Fig.~\ref{fig:fig10}(a). At
these low energies, the quasiparticles form banana-shaped regions around the
nodal points [Fig.~\ref{fig:fig10}(b)]. Increasing energy from the onset of
scattering at $\Omega_s$, the scattering rate increases and reaches its maximum
at $\Delta_0+\Omega_s$ [see Eq.~(\ref{eq:simple_damping_rate})]. Since
$\Delta_p\ll\Delta_0+\Omega_s$, the antinodal excitations at
$|\omega|=\Delta_p$ remain rather sharp [Fig.~\ref{fig:fig10}(c)]. When the
scattering rate is maximum at $|\omega|=\Delta_0+\Omega_s$, in contrast, the
excitations are very broad [Fig.~\ref{fig:fig10}(d)]. At this energy, the real
part of the self-energy changes sign, and the quasiparticle dispersion
correspondingly jumps from below to above the noninteracting dispersion. This
abrupt change in dispersion renormalization occurs \emph{at the same energy in
the whole Brillouin zone}, both for occupied and for empty states, and leads to
a removal of spectral weight which is responsible for the dip in the DOS. Near
the antinode, the $d$-wave gap induces additional structures: the minimum-gap
locus is close to---but not exactly at---the position of the noninteracting
Fermi surface along the $(\pi,0)$--$(\pi,\pi)$ line, due to Fermi-surface
renormalization. One also notices the reduction of the gap $\Delta_p$ with
respect to $\Delta_0$, due to the competition between pairing and
renormalization (Sec.~\ref{sec:model}). In the low-energy region, the
weakly dispersing quasiparticle branch near $(\pi,0)$ has lower energy than the
corresponding BCS branch, while above the dip energy, the quasiparticle energy
is higher than the noninteracting and BCS dispersions, like in the nodal region.

For a more quantitative comparison of our results with real ARPES data, we have
simulated the upper panels of Fig.~2 in Ref.~\onlinecite{Ideta-2008}. We set the
temperature to 10~K, and filter our spectral function with a Gaussian
representing an energy resolution of 18~meV and a momentum resolution of
$0.05/a$ \cite{Ideta-2008}. The result is displayed in Fig.~\ref{fig:fig11}.
Close to the nodal direction (left and middle panels), the agreement is good.
The model is too crude to completely capture the measured dispersion in the
region of the dip: the simulated dispersion jumps from below to above the
noninteracting dispersion, while the experiment interpolates smoothly across the
jump. We attribute this discrepancy to additional scattering mechanisms not
included in the model, in particular those involving the continuum of spin
fluctuations \cite{Eschrig-2003}.

The energy of the ``kink'' around $-70$~meV is well reproduced by the
calculation. We emphasize that the feature corresponding to the energy
$\Delta_0+\Omega_s$ is \emph{not} the kink, but the midpoint of the jump, where
the quasiparticle dispersion crosses the noninteracting dispersion. It is also
worth stressing that this energy does not disperse, and is given by
$\Delta_0+\Omega_s$ at the node, in spite of the fact that the gap vanishes, as
demonstrated in Fig.~\ref{fig:fig10}(a). The experimental determination of this
energy scale by ARPES requires an assumption for the noninteracting dispersion.
Our results call into question the assumption made in
Ref.~\onlinecite{Ideta-2008}, that the quasiparticle and noninteracting
dispersions meet near $-200$~meV. This assumption has direct implications for
the value of the self-energy deduced from ARPES. In particular, the real part
vanishes where the quasiparticle and noninteracting dispersions are equal, i.e.,
near $-200$~meV in Ref.~\onlinecite{Ideta-2013}. Using the dispersion from our
fits, the real part of the self-energy would vanish at the energy
$-(\Delta_0+\Omega_s)$, which is consistent with a maximum of scattering rate at
this energy. In the antinodal region (right panel of Fig.~\ref{fig:fig11}),
there are differences between our results and the ARPES data: the low-energy
part, below the kink, is too dispersive in the model, while the high-energy part
is not dispersive enough. This could be partly due to different Fermi surfaces
in the experiment and in the model, which imply that the segments of the
dispersion considered in both are not exactly identical. It could also be the
consequence of scattering processes neglected in the model. Despite these
differences, the kink energy and the minimum in the spectral intensity, between
the low-energy and high-energy parts, are very similar. The overall agreement
between Fig.~\ref{fig:fig11} and Fig.~2 of Ref.~\onlinecite{Ideta-2008} supports
the claim that STM tunneling spectra, although they come from a
momentum-integrating probe, do contain the necessary information needed to
reconstruct the low-energy momentum-resolved spectral function.

\section{Summary and conclusions}
\label{sec:conclusion}

We have performed an analysis of STM spectra measured on optimally doped
Bi-2223, by means of a strong-coupling model which takes into account the Van
Hove singularity within a one-band description, a BCS gap with pure $d$-wave
symmetry, and a coupling to the $(\pi,\pi)$ spin resonance. This model can
reproduce the experimental spectra (Fig.~\ref{fig:fig06}), with values of the
parameters which are sound, and consistent with values obtained by other
experimental probes. The inhomogeneity of the electronic properties on the
sample surface allowed us to study variations in parameters as a function of the
spectral gap $\Delta_p$. Assuming a one-to-one correspondence between $\Delta_p$
and the hole doping level, changes in $\Delta_p$ may be interpreted as local
variations of doping. The main trends are that the Van Hove singularity moves to
lower energies with increasing $\Delta_p$, and the energy of the spin resonance
decreases (Fig.~\ref{fig:fig08}). The former supports the interpretation that
larger gaps correspond to lower doping, and the latter supports the claim that
the dip is caused by the spin resonance, rather than optical phonons. The
strength of the coupling to the spin resonance, measured by the dispersion
renormalization, increases steadily for gaps larger than 42~meV, namely towards
the underdoped region of the phase diagram.

The presence of a Van Hove singularity, breaking the electron-hole symmetry of
the electronic spectrum, is unmistakable in the raw data (Fig.~\ref{fig:fig07}).
Nevertheless, the precise determination of the hopping amplitudes $t_i$ remains
a challenge, as illustrated by the large uncertainties attached to these
parameters in Table~\ref{tab:tab1}. The reason is that relatively large
variations of the $t_i$'s can collaborate to induce marginal changes in the
electron DOS. Our confidence in the fitted tight-binding dispersion stems from
the ability of the whole model to reproduce momentum-resolved ARPES data with
good accuracy. In turn, our determination of the band structure may provide
indications on how to extract the self-energy from ARPES measurements.

\acknowledgments

We acknowledge useful discussions with M. Eschrig. This work was supported by
the Swiss National Science Foundation through Division II and MaNEP.

\appendix*

\section{\boldmath The function $B(\omega,E)$}

Equations~(\ref{eq:Sigma}) to (\ref{eq:A}) imply that the function $B(\omega,E)$
entering Eq.~(\ref{eq:Sigma1}) reads
    \begin{multline}\label{eq:B}
        B(\omega,E)=\Lambda^2\int_{-\infty}^{\infty} d\varepsilon_1d\varepsilon_2\,
        [L_{\Gamma_s}(\varepsilon_1-\Omega_s)-L_{\Gamma_s}(\varepsilon_1+\Omega_s)]\\
        \times L_{\Gamma}(\varepsilon_2-E)\,
        \frac{b(\varepsilon_1)+f(-\varepsilon_2)}{\omega-\varepsilon_1-\varepsilon_2+i0^+},
    \end{multline}
where $b$ and $f$ are the Bose and Fermi functions, respectively. Using the
identity $\int
dx\,L_{\Gamma}(x)/(z-x)=1/[z+i\Gamma\,\text{sign}(\text{Im}\,z)]$, we obtain
    \begin{multline}
        \Lambda^{-2}B(\omega,E)=\int_{-\infty}^{\infty}
        d\varepsilon\,\frac{L_{\Gamma_s}(\varepsilon-\Omega_s)
        b(\varepsilon)}{\omega-E+i\Gamma-\varepsilon}\\
        +\int_{-\infty}^{\infty}
        d\varepsilon\,\frac{L_{\Gamma}(\varepsilon-E)f(-\varepsilon)}
        {\omega-\Omega_s+i\Gamma_s-\varepsilon}-\big\{\Omega_s\to-\Omega_s\big\}.
    \end{multline}
The symbol in braces means that the result obtained from the first two terms on
the right-hand side must be antisymmetrized with respect to $\Omega_s$. The remaining
integrals can be evaluated by closing the integration contour in the complex
plane. The integrands can we rewritten in terms of products of simple poles by
means of the identities
    \begin{eqnarray}
        L_{\Gamma}(z)&=&\frac{1}{2\pi i}\left(\frac{1}{z-i\Gamma}-\frac{1}{z+i\Gamma}\right)\\
        b(z)&=&\frac{1}{\beta}\sum_{i\Omega_n}\frac{e^{i\Omega_n0^+}}{z-i\Omega_n}\\
        f(z)&=&-\frac{1}{\beta}\sum_{i\omega_n}\frac{e^{i\omega_n0^+}}{z-i\omega_n}.
    \end{eqnarray}
Only one-half of the poles of the Bose or Fermi function are enclosed in the
contour and give a contribution. Therefore, the integrals involve semi-infinite
sums over Matsubara frequencies, which can be converted into the digamma
function $\psi$, using the relation
    \begin{equation}
        \lim_{M\to\infty}\,\sum_{n=0}^{M}\frac{e^{\pm in0^+}}{n+z}=\ln M-\psi(z).
    \end{equation}
The final result is
    \begin{multline}
        \Lambda^{-2}B(\omega,E)=\frac{b(\Omega_s-i\Gamma_s)+\frac{1}{2\pi i}\,
        \psi\left[\frac{\beta}{2\pi i}\left(\Omega_s-i\Gamma_s\right)\right]}
        {\omega-E-\Omega_s+i(\Gamma+\Gamma_s)}\\
        +\frac{\frac{\Gamma_s}{\pi}\,\psi\left[\frac{\beta}{2\pi i}
        \left(\omega-E+i\Gamma\right)\right]}{(\omega-E-\Omega_s+i\Gamma)^2+\Gamma_s^2}
        -\frac{\frac{1}{2\pi i}\,\psi\left[\frac{\beta}{2\pi i}
        \left(\Omega_s+i\Gamma_s\right)\right]}{\omega-E-\Omega_s+i(\Gamma-\Gamma_s)}\\
        +\frac{f(-E+i\Gamma)+\frac{1}{2\pi i}\,\psi\left[\frac{1}{2}+\frac{\beta}{2\pi i}
        \left(E-i\Gamma\right)\right]}{\omega-E-\Omega_s+i(\Gamma+\Gamma_s)}\\
        +\frac{\frac{\Gamma}{\pi}\,\psi\left[\!\frac{1}{2}+\frac{\beta}{2\pi i}
        \left(\omega-\Omega_s+i\Gamma_s\right)\!\right]}
        {(\omega-E-\Omega_s+i\Gamma_s)^2+\Gamma^2}
        -\frac{\frac{1}{2\pi i}\,\psi\left[\!\frac{1}{2}+\frac{\beta}{2\pi i}
        \left(E+i\Gamma\right)\!\right]}{\omega-E-\Omega_s+i(\Gamma_s-\Gamma)}\\
        -\big\{\Omega_s\to-\Omega_s\big\}.
    \end{multline}
The function $B$ simplifies considerably in the case of a sharp resonance
($\Gamma_s=0^+$) and for sharp Bogoliubov quasiparticles ($\Gamma=0^+$), as well
as zero temperature. In this case, we have
    \begin{equation}\label{eq:Bsimple}
        B(\omega,E)=\frac{\Lambda^2}{\omega-E-\Omega_s\,\text{sign}(E)+i0^+},
    \end{equation}
as can be readily deduced from Eq.~(\ref{eq:B}), by replacing the Lorentzians by
delta functions.

\bibliography{bib,spot,plus}

\end{document}